\documentclass[twocolumn]{aastex701}
\usepackage{url}

\begin{document}

\title{The Structure and Kinematics of Three Class 0 Protostellar Jets from JWST}


\author[0000-0002-6136-5578]{Samuel A. Federman}
\affiliation{Ritter Astrophysical Research Center, Dept. of Physics and Astronomy, University of Toledo, Toledo, OH, US}
\affiliation{INAF-Osservatorio Astronomico di Capodimonte, IT}
\email{sam.federman@gmail.com}

\author[0000-0001-7629-3573]{S. Thomas Megeath}
\affiliation{Ritter Astrophysical Research Center, Dept. of Physics and Astronomy, University of Toledo, Toledo, OH, US}
\email{tommegeath@gmail.com}

\author[0000-0001-8876-6614]{Alessio Caratti o Garatti}
\affiliation{INAF-Osservatorio Astronomico di Capodimonte, IT}
\email{alessio.caratti@inaf.it}

\author[0000-0002-0554-1151]{Mayank Narang}
\affiliation{Jet Propulsion Laboratory, California Institute of Technology, 4800 Oak Grove Drive, Pasadena, CA 91109, USA}
\email{mayank.narang@jpl.nasa.gov}

\author[0000-0002-9497-8856]{Himanshu Tyagi}
\affiliation{Tata Institute of Fundamental Research, Mumbai, Maharashtra, IN}
\email{himanshu.tyagi@tifr.res.in}

\author[0000-0001-5175-1777]{Neal J. Evans II}
\affiliation{Department of Astronomy, The University of Texas at Austin, 2515 Speedway, Stop C1400, Austin, Texas 78712-1205, USA}
\email{nje@astro.as.utexas.edu}

\author[0000-0001-9071-1508]{Carolin N. Kimmig}
\affiliation{Dipartimento di Fisica, Università degli Studi di Milano, via Celoria
16, 20133 Milano, Italy}
\email{carolin.kimmig@unimi.it}

\author[0000-0002-9470-2358]{Łukasz Tychoniec}
\affiliation{Leiden Observatory, Leiden University, PO Box 9513, 2300 RA, Leiden, The Netherlands}
\email{tychoniec@strw.leidenuniv.nl}

\author[0000-0002-1700-090X]{Henrik Beuther}
\affiliation{Max Planck Institute for Astronomy, Heidelberg, Baden Wuerttemberg, DE}
\email{beuther@mpia-hd.mpg.de}

\author[0000-0003-2300-8200]{Amelia Stutz}
\affiliation{Departamento de Astronomía, Universidad de Concepción, Casilla 160-C, Concepción, Chile}
\email{amelia.stutz@gmail.com}

\author[0000-0002-3530-304X]{P. Manoj}
\affiliation{Tata Institute of Fundamental Research, Mumbai, Maharashtra, IN}
\email{mpuravankara@gmail.com}

\author[0000-0002-6447-899X]{Robert Gutermuth}
\affiliation{University of Massachusetts Amherst, Amherst, MA, US}
\email{rob.gutermuth@gmail.com}

\author[0000-0001-7491-0048]{Tyler L. Bourke}
\affiliation{SKA Observatory, Jodrell Bank, Lower Withington, Macclesfield SK11 9FT, UK}
\email{tyler.bourke@skao.int}

\author[0000-0003-1665-5709]{Joel Green}
\affiliation{Space Telescope Science Institute, 3700 San Martin Drive, Baltimore, MD 21218, US}
\email{jgreen@stsci.edu}

\author[0000-0003-1430-8519]{Lee Hartmann}
\affiliation{University of Michigan, Ann Arbor, MI, US}
\email{lhartm@umich.edu}

\author[0000-0001-9443-0463]{Pamela Klaassen}
\affiliation{United Kingdom Astronomy Technology Centre, Edinburgh, GB}
\email{pamela.klaassen@stfc.ac.uk}

\author[0000-0003-2309-8963]{Rolf Kuiper}
\affiliation{Faculty of Physics, University of Duisburg-Essen, Lotharstra{\ss}e 1, D-47057 Duisburg, Germany}
\email{rolf.kuiper@uni-due.de}

\author[0000-0002-4540-6587]{Leslie W. Looney}
\affiliation{Department of Astronomy, University of Illinois, 1002 West Green St, Urbana, IL 61801, US}
\email{lwl@illinois.edu}

\author[0000-0002-4448-3871]{Pooneh Nazari}
\affiliation{Leiden Observatory, Universiteit Leiden, Leiden, Zuid-Holland, NL}
\email{Pooneh.Nazari@eso.org}


\author[0000-0002-5812-9232]{Thomas Stanke}
\affiliation{Max-Planck Institut f\"ur Extraterrestrische Physik, Garching bei München, DE}
\email{tstanke049@gmail.com}

\author[0000-0001-8302-0530]{Dan M. Watson}
\affiliation{University of Rochester, Rochester, NY, US}
\email{dmw@pas.rochester.edu}

\author[0000-0001-8227-2816]{Yao-Lun Yang}
\affiliation{Star and Planet Formation Laboratory, RIKEN Pioneering Research Institute, Wako-shi, Saitama, 351-0106, Japan}
\email{yaolunyang.astro@gmail.com}

\author[0000-0001-9030-1832]{Wafa Zakri}
\affiliation{Physical Sciences Department, Jazan University, Jazan, Saudi Arabia}
\email{wafa.zakri@rockets.utoledo.edu}


\begin{abstract}
We present observations of jets within 2000~au of three deeply embedded protostars using 2.9-27~$\mu$m observations with JWST. These observations show the morphologies and kinematics of the collimated jets from three protostars, the low-mass Class~0 protostars B335 and HOPS~153, and the intermediate-mass protostar HOPS~370. These jets are traced by shock-ionized fine-structure line emission observed with the JWST NIRSpec and MIRI IFUs. We find that [Fe~II] emission traces the full extent of the inner 1000 to 2000~au of the jets, depending on distance to the protostar, while other ions mostly trace isolated shocked knots. The jets show evidence of wiggling motion in the plane of the sky as well as asymmetries between blue and red-shifted lobes. The widths of the jets increase non-monotonically  with distance from the central protostar, with opening angles ranging from 2.1$^{\circ}$ to $<10.1^{\circ}$ for the three protostars in the sample. The jets have total velocities ranging from 147 to 184 km~s$^{-1}$ after correcting for disk inclination. For B335, an 8-month gap between NIRSpec and MIRI MRS observations enabled measurement of the tangential velocity of a shocked knot; in combination with the radial velocity, this shows that the jet has a different inclination than the outflow cavity. We find multiple knots before and during a recent outburst in B335, although the knots were more frequent during the burst. The asymmetries between blue- and red-shifted lobes strongly suggest complex interactions between the circumstellar disks and magnetic fields.

\end{abstract}

\keywords{}

\section{Introduction}

High-velocity collimated jets are a natural consequence of accreting, rotating systems with accretion mediated by magnetism \citep{Blanford_1982, Frank_2014, Bally_2016}. Protostars are no exception \citep[see][and references therein]{lee_2020, Ray_2021}, driving powerful jets that transport angular momentum out of the system, expelling mass and kinetic energy into the surrounding environment \citep[e.g.][]{Frank_2014,Bally_2016},  and allowing matter to accrete onto the growing protostar while simultaneously dispersing the infalling material. It has been demonstrated that collimated jets traced by shocked [Fe~II] emission detected with the NIRSpec IFU on JWST are a common feature of young, envelope-dominated Class~0 protostars across the mass spectrum \citep{Federman_2024, Narang_2024, Caratti_2024}. Several additional [Fe~II] and other ionic lines are accessible with the MIRI MRS IFU covering wavelengths from 4.9-27.9 $\mu$m (R $\sim{2000-4000}$), which provides an opportunity to analyze the morphologies and kinematics of the collimated jets in detail \citep[e.g.][]{Yang_2022, Barsony_2024, Nisini_2024, Tychoniec_2024, Narang_2024, Narang_2025, Okoda_2025, vanDishoeck_2025}.

Investigating Protostellar Accretion and Outflow Across the Mass Spectrum (IPA) is a Cycle 1 JWST GO program (PID:1802) that targeted five young, embedded protostars in their primary accretion phase, spanning the protostellar mass spectrum. The three low-mass protostars, I16253-2429, B335, and HOPS~153, have Class 0 SEDs and are shown to be envelope-dominated \citep{Stanke_2006, Chandler_1990, Furlan_2016, Federman_2023}. The intermediate-mass protostar, HOPS~370, has an SED at the boundary between Class 0 and Class I, and is also envelope-dominated \citep{Tobin_2020B, Furlan_2016, Federman_2023}. These all appear to be protostars in their primary mass assembly phase, when most of a protostar's mass is accreted \citep{Federman_2023}. 

\citet{Federman_2024} provided an overview of the IPA NIRSpec data, in which they found that shock-ionized [Fe~II] is an ubiquitous tracer of collimated jets from Class~0 protostars, with shocked knots traced by molecular, atomic, and ionic emission. They also found collimated molecular emission in the jet of the intermediate-mass protostar HOPS~370. More commonly, however, the warm H$_{2}$ traces molecular emission extending through the cavities, including inner shells of material that may be driven by jet knots. These jets appear to be the high-velocity centers of nested outflows, where shock velocities are high enough to ionize the neutral gas. The lack of molecular emission is likely due to dissociation in the jet and/or depletion of molecules in the inner regions of disks from which these jets are launched. Thus, these high-velocity protostellar jet components must be studied in ionic tracers.

\citet{Narang_2024} investigated the nature of the collimated jet primarily traced by the [Fe~II] emission for the lowest-mass protostar in the IPA sample, IRAS~16253-2429 (hereafter IRAS~16253). The authors used a combination of the higher spatial resolution NIRSpec data and lower spatial resolution MIRI data to evaluate the morphology of the jet. With the higher spectral resolution of the MIRI data, \citet{Narang_2024} resolved the radial velocity structure along the jet. They found that the jet velocites remained constant along the length of the observed jet. Using shock models to estimate the shock densities and velocities, they determined that IRAS~16253 is currently in a quiescent phase with a low mass-outflow rate and low mass-accretion rate. This required that IRAS~16253 must have been in a state of higher mass accretion in the past. 

In this paper, we present the jet morphologies and kinematics for the two low-mass protostars, B335 (0.25~M$_{\odot}$, Table~\ref{tab:sources}) and HOPS~153 (0.6~M$_{\odot}$), and the intermediate-mass protostar HOPS~370 (2.5~M$_{\odot}$), from the IPA sample. Here we focus on tracers of the jet with the MIRI MRS IFU in ionic lines of iron, nickel, neon, and argon. These MIRI data present a more rigorous examination of the ionic emission in jets than is possible with NIRSpec, for which the lines are often weak or contaminated. In addition, they provide information on the kinematics of the gas. This paper will focus on the structure and kinematics of the jets. A companion paper will use shock modeling of the ionic lines mapped out in this paper to determine shock velocities and pre-shock densities, and thereby measure the flows of mass, momentum, and energy in the jets.



\section{Sample and Observations}

The analyzed protostars are from the IPA sample described in \citet{Federman_2024}, which targeted 5 Class~0 protostars spanning a roughly logarithmic range in protostellar mass and luminosity with intermediate to high inclinations to facilitate observations of the extended outflows. In this analysis we exclude the lowest-mass protostar IRAS~16253, which was analyzed in \citet{Narang_2024}, and the high-mass protostar IRAS~20126+4104. We exclude the high-mass protostar because it has a much more complicated jet morphology than the lower-mass protostars, along with other significant variations as detailed in \citet{Federman_2024}, and will be analyzed separately. The main properties of the three protostars are listed in Table~1 of \citet{Federman_2024}, and the NIRSpec observations are described in the same paper; the relevant properties for this work are shown in Table~\ref{tab:sources}. Note that in \citet{Federman_2024}, the $<$10~au disk radius quoted for B335 came from \citet{Yen_2015} based on a distance of 100~pc, but in this work we adopt the more recent value of $<$16~au quoted in \cite{Evans_2023} based on the 165~pc distance found in \citet{Watson_2020}. We also updated the jet position angle of HOPS~370 from 5$^{\circ}$ in \citet{Federman_2024} to 6$^{\circ}$, based on visual inspection.

\begin{deluxetable*}{cccccc}[t]\label{tab:sources}
\tablecaption{Analyzed source properties.}
\centering
\tablehead{\colhead{\textbf{Source:}} & \colhead{B335} & \colhead{HOPS 153} & \colhead{HOPS 370} & \colhead{References}}
\startdata
    \textbf{RA:} & 19:37:00.89 & 05:37:57.021 & 05:35:27.635 & 1,2,3 \\
    \textbf{Dec:} & +07:34:09.6 & -07:06:56.25 & -05:09:34.44 & 1,2,3\\
    \textbf{Distance:} & 165~pc & 390~pc & 390~pc & 4,2,2 \\
    \textbf{L$_{bol}$:} & 1.4~L$_{\odot}$ & 3.8~L$_{\odot}$ & 315.7~L$_{\odot}$ & 5,2,2 \\
    \textbf{Stellar Mass:} & 0.25~M$_{\odot}$ & 0.6~M$_{\odot}$ & 2.5~M$_{\odot}$ & 5,3,10 \\
    \textbf{Disk Radius:} & $<$16~au & 150~au & 100~au & 5,2,2 \\
    \textbf{Disk Inclination:} & 83.5$^{\circ}$* & 74.5$^{\circ}$ & 72.2$^{\circ}$ & 6/7,2,2 \\
    \textbf{Disk Major Axis Position Angle$\ast\ast$:} & 5$^{\circ}$ & 33$^{\circ}$ & 110$^{\circ}$ & 8,2,2 \\
    \textbf{Jet Position Angle$\ast\ast$:} & 90$^{\circ}$ & -55$^{\circ}$ & 6$^{\circ}$ & 9,9, this work
\enddata
\tablecomments{References: (1) \citealt{Maury_2018}, (2) \citealt{Tobin_2020A}a, (3) \citealt{Tobin_2020B}b, (4) \citealt{Watson_2020}, (5) \citealt{Evans_2023}, (6) \citealt{Hirano_1988}, (7) \citealt{Stutz_2008}, (8) \citealt{Bjerkeli_2019}, (9) \citealt{Federman_2024}, (10) J. J. Tobin 2024, private communication \\
$\ast$The disk inclination of B335 is the average of the 87$^{\circ}$ inclination quoted from \citet{Stutz_2008} and the 80$^{\circ}$ inclination from \citep{Hirano_1988}.}
$\ast\ast$ Disk and jet position angles are measured relative to East of North.
\end{deluxetable*}

The MIRI MRS observations of the three protostars were taken between 2023 March 12 and May 12. The targets were observed in all four MIRI MRS channels covering 4.9-27.9 $\mu$m with a spectral resolving power ranging from R$\sim$2000-4000 \citep{Pontoppidan_2024b, Banzatti_2025} and angular resolution ranging from 0\farcs27 at the shortest wavelength to 1'' at the longest wavelength \citep{Law_2023}. The FOV of the MIRI MRS IFU varies from $3\farcs2 \times 3\farcs7$ in the shortest wavelength subchannel of Channel 1 to $6\farcs6 \times 7\farcs7$ in the longest wavelength subchannel of Channel 4. The IPA MIRI observations covered a 2x2 mosaic with 10\% overlap and a 4-point dither pattern, similar to the NIRSpec observations. The exposures had 12 groups in one integration using the SLOWR1 readout mode for B335 and HOPS~153, and 50 groups in two integrations using the FASTR1 readout mode for the brighter HOPS~370. Background observations were taken with the same configurations for each target, in a location selected to avoid stellar contamination and bright nebulosity. Bonus parallel imaging was obtained in the F770W, F1280W, and F1500W filters at both the target and background positions for each protostar. The imaging exposures had 25 groups in four integrations using the FASTR1 readout mode in each filter for B335 and HOPS~153, and 25 groups in one integration using the FASTR1 readout mode in each filter for HOPS~370. The MIRI data were reduced using the JWST pipeline version 11.17.26 and the JWST Calibration References Data System context version jwst\_1276.pmap. Refer to \citet{Narang_2024} for the details of the customization of the pipeline reduction for the IPA MIRI data \citep[see also][]{Neufeld_2024, Tyagi_2025}. For this work, we use data in which the background-subtraction step in the pipeline was enabled, to account for possible contamination from the larger molecular cloud. The simultaneous imaging fields were used for astrometric calibration of the NIRSpec and MIRI data by matching sources in the MIRI imaging field to Spitzer sources with an astrometry measured in Gaia as described in \citet{Federman_2024}. 

\section{Results}\label{sec:iron_jet_results}

In this section, we present the results of the analysis of the collimated jets of B335, HOPS~153, and HOPS~370 through shock-ionized fine-structure line emission. We detail the morphology of the jets observed in [Fe~II] through their knots, widths, wiggles, bends, and deviations from bipolar symmetry between the blue and red-shifted lobes. We then investigate the radial velocity structure, $V_{rad}$, along the [Fe~II] jets. The [Fe~II] lines used in the analysis include the 4.115 $\mu$m line from NIRSpec and the 5.340, 17.936, 24.519, and 25.988 $\mu$m lines from MIRI. We also create line maps of ionic emission from [Ni~II] (6.637~$\mu$m), [Ar~II] (6.985~$\mu$m), and [Ne~II] (12.814~$\mu$m), which prominently trace shocked knots, and in some cases, the extended jet. The angular, spatial, and velocity resolution of each line are listed in Table~\ref{tab:resolution}. While these are the native resolution of the data, with Gaussian fitting we can achieve more precise measurements with the standard deviation of the fits being inversely proportional to the signal-to-noise ratio \citep[e.g.][]{Landman_1982, Hatzes_1992}. In Appendix~\ref{sec:gauss_precision}, we list the average angular/spatial and velocity precision achievable with Gaussian fitting for each of the [Fe~II] lines used in the analysis (Table~\ref{tab:gauss_precision}). We generate continuum-subtracted cubes by subtracting a linear fit to the local continuum around each line at each pixel, and line maps were made by summing over the full-width at zero intensity line width.

\begin{table*}[t]\centering
\caption{Angular and spatial resolution at the distance of B335 and the Orion protostars for all the ionic lines used in the analysis, and the spectral resolving power and velocity resolution of each line.}
\begin{tabular}{ccccc}
    Line & $\theta$ & Spatial Resolution* & R & Velocity Resolution \\ \hline \hline
    [Fe~II] 4.115~$\mu$m & 0.19'' & 31~au/74~au & 1000 & 300~km~s$^{-1}$ \\ \hline
    [Fe~II] 5.340~$\mu$m & 0.28'' & 47~au/110~au & 3543 & 85~km~s$^{-1}$ \\ \hline
    [Fe~II] 17.936~$\mu$m & 0.70'' & 115~au/272~au & 3356 & 89~km~s$^{-1}$ \\ \hline
    [Fe~II] 24.519~$\mu$m & 0.92'' & 151~au/357~au & 1845 & 163~km~s$^{-1}$ \\ \hline
    [Fe~II] 25.988~$\mu$m & 0.96'' & 159~au/376~au & 2162 & 139~km~s$^{-1}$ \\ \hline
    [Ni~II] 6.637~$\mu$m & 0.33'' & 54~au/127~au & 3446 & 87~km~s$^{-1}$ \\ \hline
    [Ne~II] 12.814~$\mu$m & 0.53'' & 87~au/206~au & 2991 & 100~km~s$^{-1}$ \\ \hline
    [Ar~II] 6.985~$\mu$m & 0.34'' & 56~au/131~au & 3655 & 82~km~s$^{-1}$ \\ \hline
\end{tabular}
\label{tab:resolution}
\tablecomments{Spatial resolution values are for d = 165~pc (B335) and d = 390~pc (HOPS~153 and HOPS~370).}
\end{table*}

\subsection{Jet Morphology}\label{sec:morphology}

\begin{figure*}[t]
    \centering
    \includegraphics[width=.99\textwidth]{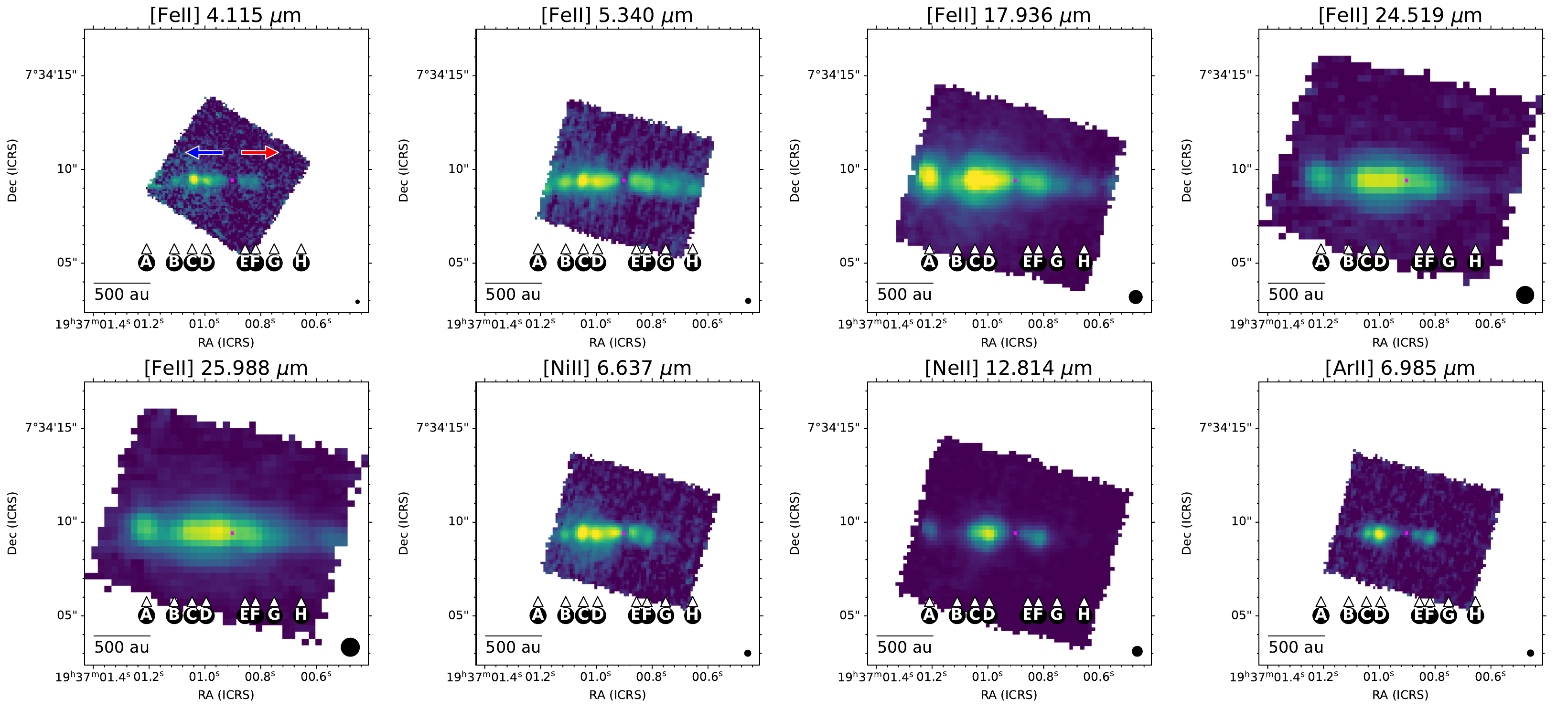}
    \caption{Fine-structure emission line maps for B335. All images are displayed with an arcsinh scale. Blue and red arrows indicate the direction of the blue- and red-shifted lobes, respectively. \textbf{Top:} The 4.115, 5.340, 17.936, and 24.519 $\mu$m [Fe~II] line maps from left to right, respectively. \textbf{Bottom:} The [Fe~II] 25.988, [Ni~II] 6.637, [Ne~II] 12.814, and [Ar~II] 6.985 $\mu$m line maps from left to right, respectively. The 4.115 $\mu$m line is from the NIRSpec data; the rest are from MIRI. The FOV and angular resolution of the MIRI data increases with wavelength; a 500~au scalebar is shown at the bottom left corner, and a beam representing the angular resolution is shown at the bottom right corner. A magenta line marks the size and position angle of the disk, centered at the ALMA position of the protostar. Because the disk of B335 has a radius of $<$16~au, the line appears as a small dot. Shocked knots are labeled along the bottom of the map with arrows pointing to the knots. The larger MIRI FOV covers an additional shocked emission knot to the east that is not covered in the NIRSpec FOV.}
    \label{fig:b335_jetmaps}
\end{figure*}

\begin{figure*}[t]
    \centering
    \includegraphics[width=.99\textwidth]{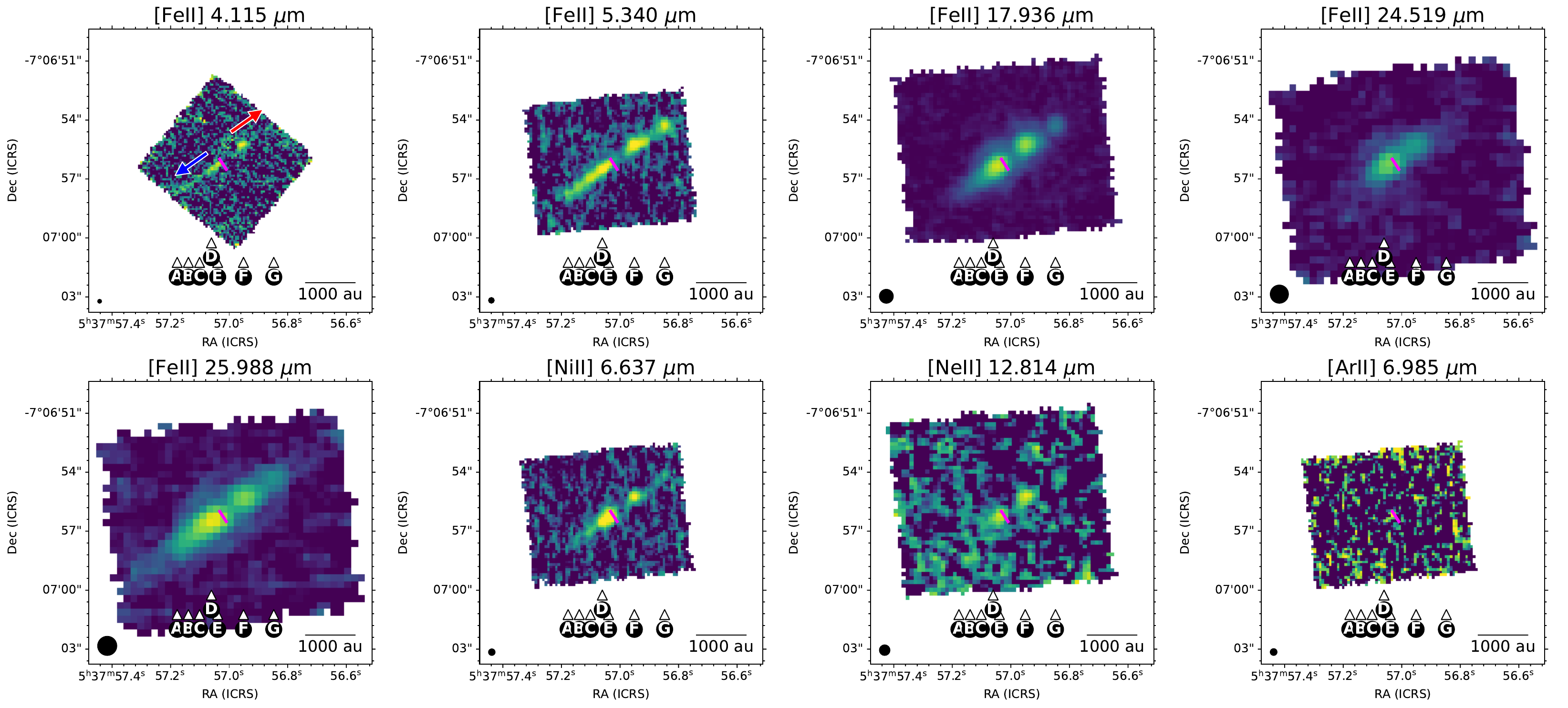}
    \caption{Fine-structure emission line maps for HOPS~153. All images are displayed with an arcsinh scale. Blue and red arrows indicate the direction of the blue- and red-shifted lobes, respectively. \textbf{Top:} The 4.115, 5.340, 17.936, and 24.519 $\mu$m [Fe~II] line maps from left to right, respectively. \textbf{Bottom:} The [Fe~II] 25.988, [Ni~II] 6.637, [Ne~II] 12.814, and [Ar~II] 6.985 $\mu$m line maps from left to right, respectively. A 1000~au scalebar is shown at the bottom right corner, and a beam representing the angular resolution is shown at the bottom left corner. A magenta line marks the size and position angle of the disk, centered at the ALMA position of the protostar. Shocked knots are labeled along the bottom of the map with arrows pointing to the knots. The larger MIRI FOV covers an additional shocked emission knot that is not covered in the NIRSpec FOV.}
    \label{fig:hops153_jetmaps}
\end{figure*}

\begin{figure*}[t]
    \centering
    \includegraphics[width=.99\textwidth]{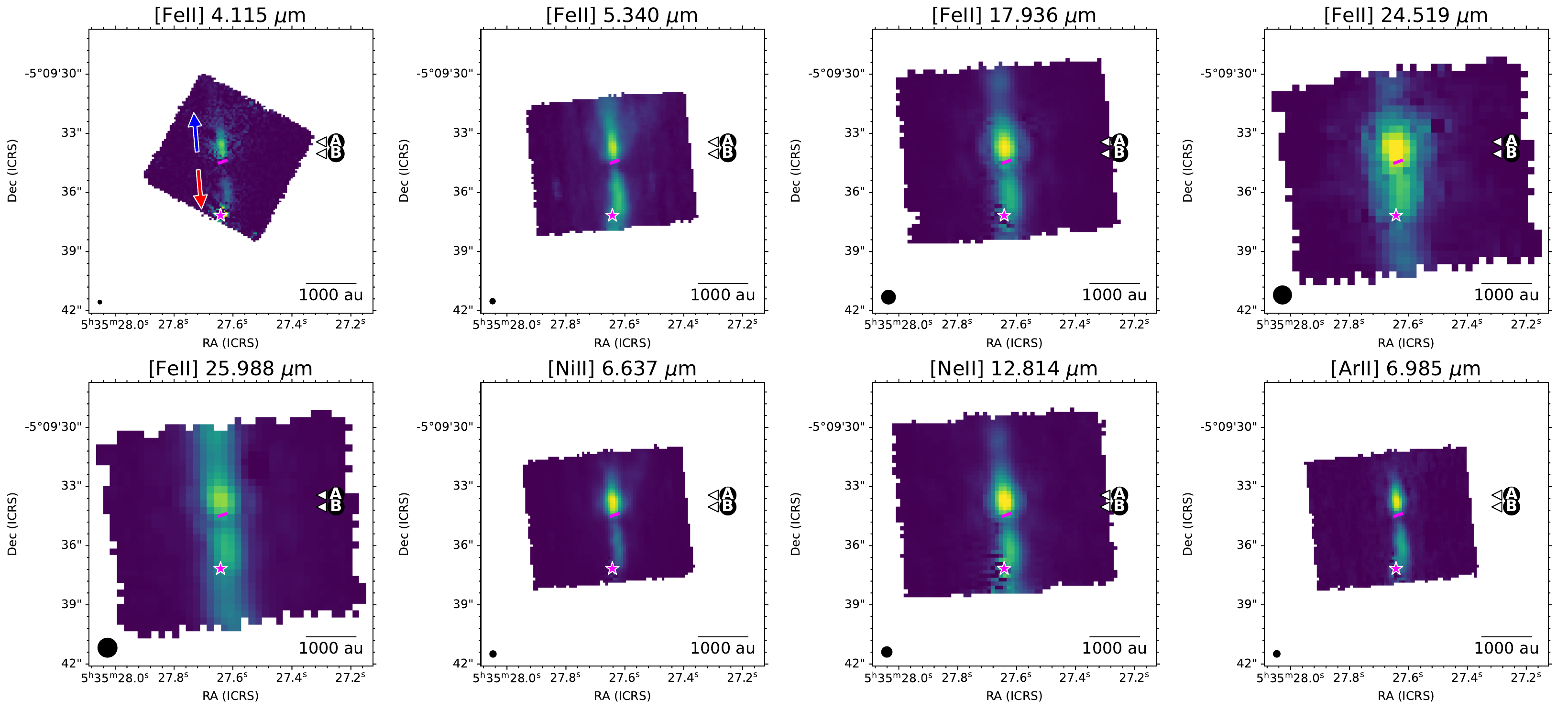}
    \caption{Fine-structure emission line maps for HOPS~370. All images are displayed with an arcsinh scale. Blue and red arrows indicate the direction of the blue- and red-shifted lobes, respectively. \textbf{Top:} The 4.115, 5.340, 17.936, and 24.519 $\mu$m [Fe~II] line maps from left to right, respectively. \textbf{Bottom:} The [Fe~II] 25.988, [Ni~II] 6.637, [Ne~II] 12.814, and [Ar~II] 6.985 $\mu$m line maps from left to right, respectively. A 1000~au scalebar is shown at the bottom right corner, and a beam representing the angular resolution is shown at the bottom left corner. A magenta line marks the size and position angle of the disk, centered at the ALMA position of the protostar. Two shocked knots are labeled along the right side of the map with arrows pointing to the knots. A magenta star marks the position of the Class~II companion to the south.}
    \label{fig:hops370_jetmaps}
\end{figure*}

In Figures~\ref{fig:b335_jetmaps}-\ref{fig:hops370_jetmaps}, we display the emission maps of the key fine-structure lines used in the analysis of B335, HOPS~153, and HOPS~370. The angular and spectral resolutions and FOVs vary with wavelength and MIRI subchannel; a scalebar is displayed in the corner of each map for reference. A magenta line marks the size and position angle of the disk centered at the position of the protostar determined by ALMA (sub)-mm continuum \citep{Bjerkeli_2019, Tobin_2020A, Federman_2024}.

\subsubsection{Knots of Shock-Ionized Emission}\label{sec:knots}

We identified a number of shocked knots by eye in the ionic line maps displayed in Figures~\ref{fig:b335_jetmaps}-\ref{fig:hops370_jetmaps}. In Table~\ref{tab:knot_pos}, we list the positions of the knots and the line maps in which they are observed. These positions were determined using 2D Gaussian fits within regions around the knots in the [Fe~II] 5.340~$\mu$m maps, except for knots that fall outside the field of view at that wavelength for which we use the 17.936~$\mu$m line (Appendix~\ref{sec:knot_regions}). The FWHM sizes of the knots are listed in Table~\ref{tab:knot_size} in Appendix~\ref{sec:knot_regions}. Several of these knots that fall within the NIRSpec FOV were identified in \citet{Federman_2024}; however, they have slightly different positions compared to Table~\ref{tab:knot_pos}, which are within the uncertainties in position for HOPS~153 and HOPS~370, but for B335 are larger than the uncertainties in position. \citet{Federman_2024} used the 4.889~$\mu$m [Fe~II] line to identify the shocked knots, which may suffer from contamination by the nearby CO ro-vibrational lines, affecting the measured center of the shocked knot. For B335, there were 8 months between the NIRSpec and MIRI observations, contributing to the observed difference. However, there is also a possibility that differences in angular resolution or the two lines tracing different regions of the shocked knot may contribute to the offset in positions \citep[e.g.][]{Dutta_2025}. For comparison of knot positions between NIRSpec and MIRI for this work, we adopt the 4.115~$\mu$m line, which is fainter but does not suffer from contamination by CO. 

\begin{table*}[t]
\caption{Coordinates of shocked knots identifed in the [Fe~II] MIRI MRS data, from 2D Gaussian fits. Uncertainties in RA, Dec include the uncertainties from the astrometric correction described in \citet{Federman_2024}. Distances are calculated from knot center to the position of the protostar listed in Table~\ref{tab:sources}. Positive distances correspond to blue-shifted knots, and negative distances to red-shifted knots. [Ne~II] refers to the 12.814~$\mu$m line, [Ni~II] refers to the 6.637~$\mu$m line, and [Ar~II] refers to the 6.985~$\mu$m line. Numbers in parentheses in the last column list the individual [Fe~II] wavelengths in which each knot is observed.}
\centering
\begin{tabular}{cccc}
    Label & Coordinate (RA,Dec) & Distance & Observed\\ \hline \hline
    \multicolumn{4}{c}{\textbf{B335}} \\
    A & ($294.255038^{\circ}~\pm$~0\farcs04, $7.569329^{\circ}~\pm$~0\farcs05) & 761~au~$\pm$~11~au & [Fe~II] (17.9,24.5,26.0); [Ne~II] \\ \hline
    B & ($294.254624^{\circ}~\pm$~0\farcs03, $7.569255^{\circ}~\pm$~0\farcs02) & 515~au~$\pm$~6~au & [Fe~II] (4.1,5.3,17.9,24.5,26.0); [Ni~II] \\ \hline
    C & ($294.254359^{\circ}~\pm~$0\farcs02, $7.569281^{\circ}~\pm~$0\farcs02) & 357~au~$\pm$~5~au & [Fe~II] (4.1,5.3,17.9,24.5,26.0); [Ni~II]; [Ne~II]; [Ar~II] \\ \hline
    D & ($294.254147^{\circ}~\pm~$0\farcs02, $7.569268^{\circ}~\pm~$0\farcs02) & 231~au~$\pm$~6~au & [Fe~II] (4.1,5.3,17.9,24.5,26.0); [Ni~II]; [Ne~II]; [Ar~II] \\ \hline
    E & ($294.253567^{\circ}~\pm~$0\farcs02, $7.569275^{\circ}~\pm~$0\farcs02) & -113~au~$\pm$~6~au & [Fe~II] (4.1,5.3,17.9,24.5,26.0); [Ni~II]; [Ne~II]; [Ar~II] \\ \hline
    F & ($294.253409^{\circ}~\pm~$0\farcs02, $7.569238^{\circ}~\pm~$0\farcs02) & -209~au~$\pm$~4~au & [Fe~II] (4.1,5.3,17.9,24.5,26.0); [Ni~II]; [Ne~II]; [Ar~II] \\ \hline
    G & ($294.253132^{\circ}~\pm~$0\farcs02, $7.569192^{\circ}~\pm~$0\farcs02) & -376~au~$\pm$~5~au & [Fe~II] (5.3,17.9,26.0); [Ni~II]  \\ \hline
    H & ($294.252728^{\circ}~\pm~$0\farcs02, $7.569148^{\circ}~\pm~$0\farcs02) & -617~au~$\pm$~5~au & [Fe~II] (5.3,17.9)  \\ \hline \hline
    \multicolumn{4}{c}{\textbf{HOPS~153}} \\ 
    A & ($84.488239^{\circ}~\pm$~0\farcs04, $-7.116029^{\circ}\pm$~0\farcs05) & 1076~au~$\pm$~25~au  & [Fe~II] (5.3,17.9,26.0) \\ 
    B & ($84.488080^{\circ}~\pm$~0\farcs03, $-7.115943^{\circ}~\pm$~0\farcs03) & 823~au~$\pm$~18~au & [Fe~II] (5.3,17.9,26.0) \\
    C & ($84.487921^{\circ}~\pm$~0\farcs03, $-7.115818^{\circ}~\pm$~0\farcs02) & 541~au~$\pm$~14~au & [Fe~II] (5.3,17.9,24.5,26.0); [Ni~II] \\
    D & ($84.487752^{\circ}~\pm$~0\farcs03, $-7.115695^{\circ}\pm$~0\farcs02) & 252~au~$\pm$~13~au & [Fe~II] (5.3,17.9,24.5,26.0); [Ni~II] \\
    E & ($84.487663^{\circ}~\pm$~0\farcs04, $-7.115629^{\circ}~\pm$~0\farcs03) & 106~au~$\pm$~20~au & [Fe~II] (4.1,5.3,17.9,24.5,26.0); [Ni~II]; [Ne~II]; [Ar~II] \\
    F & ($84.487296^{\circ}~\pm$~0\farcs03, $-7.115349^{\circ}\pm$~0\farcs03) & -564~au~$\pm$~14~au & [Fe~II] (4.1,5.3,17.9,24.5,26.0); [Ni~II]; [Ne~II]; [Ar~II] \\
    G & ($84.486865^{\circ}~\pm$~0\farcs01, $-7.115076^{\circ}~\pm$~0\farcs02) & -1274~au~$\pm$~8~au & [Fe~II] (5.3,17.9,26.0) \\ \hline \hline
    \multicolumn{4}{c}{\textbf{HOPS~370}} \\
    A & ($83.86518^{\circ}~\pm$~0\farcs03, $-5.159315^{\circ}~\pm$~0\farcs05) & 357~au~$\pm$~22~au & [Fe~II] (4.1,5.3,17.9,24.5,26.0); [Ni~II]; [Ne~II]; [Ar~II] \\
    B & ($83.86516^{\circ}~\pm$~0\farcs04, $-5.159426^{\circ}~\pm$~0\farcs05) & 198~au~$\pm$~24~au & [Fe~II] (4.1,5.3,17.9,24.5,26.0); [Ni~II]; [Ne~II]; [Ar~II] \\
\end{tabular}
\label{tab:knot_pos}
\end{table*}

B335 and HOPS~153 both have chains of knots along the extent of their jets, in contrast to HOPS~370 which only has two knots identified at the base of the blue-shifted lobe. For HOPS~370, the 4.115~$\mu$m map and the NIRSpec data reviewed in \citet{Federman_2024} provided additional context in identifying the two distinct knots. The rest of the HOPS~370 jet has an overall smoother distribution compared to the lower-mass B335 and HOPS~153 jets. The knots are all resolved or marginally resolved, with average sizes of 0\farcs69 (114~au), 0\farcs36 (140~au), and 0\farcs40 (156~au) for B335, HOPS~153, and HOPS~370, respectively. Because their sizes are similar to the angular resolution of the data (0\farcs28 at 5.340~$\mu$m), we subtract out the angular resolution in quadrature (i.e. $FWHM = \sqrt{FWHM_{fit}^{2}-FWHM_{PSF}^{2}}$); this results in average sizes of 0\farcs63 (104~au), 0\farcs23 (90~au), and 0\farcs29 (113~au) for B335, HOPS~153, and HOPS~370, respectively. Thus, the observed knots in the [Fe~II] jets in our data have consistent sizes of $\sim$100~au, independent of distance or protostellar mass or luminosity. These sizes are larger than the $\sim$5-20~au SiO knots noted in higher resolution 0\farcs02 (8~au) ALMA data of HH~211 \citep{Lee_2017}, but are consistent with VLT observations of H$_{2}$ shocks in the OMC region with 0\farcs08 - 0\farcs15 (31-70~au) resolution \citep{Lacombe_2004, Kristensen_2008}, which have been reproduced with C-type bow-shock models \citep[e.g.][]{Kristensen_2007, Kristensen_2008, Gustafsson_2010}. We note that one knot in our sample, knot B335-A, has a structure connecting it to H$_{2}$ emission which resembles a bow-shock \citep{Federman_2024}.

\subsubsection{Wiggles and Bends Between Blue and Red-shifted Lobes}\label{sec:wiggles}

\begin{figure*}[t]
    \centering
    \includegraphics[width=.3\textwidth]{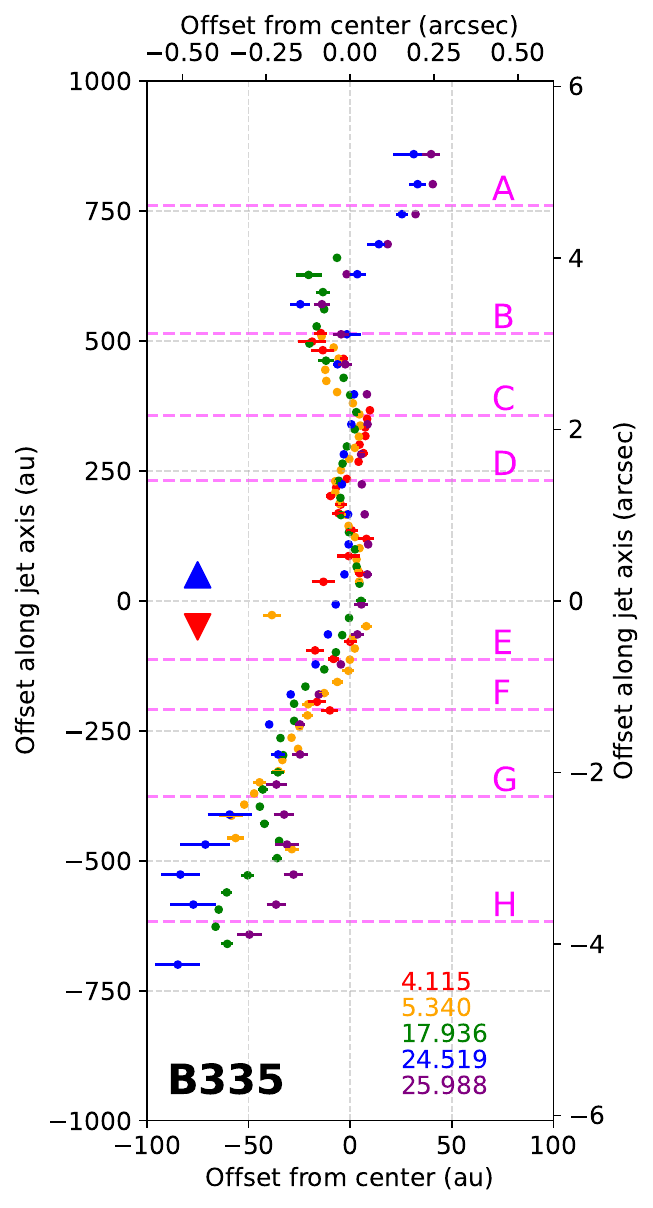}
    \includegraphics[width=.295\textwidth]{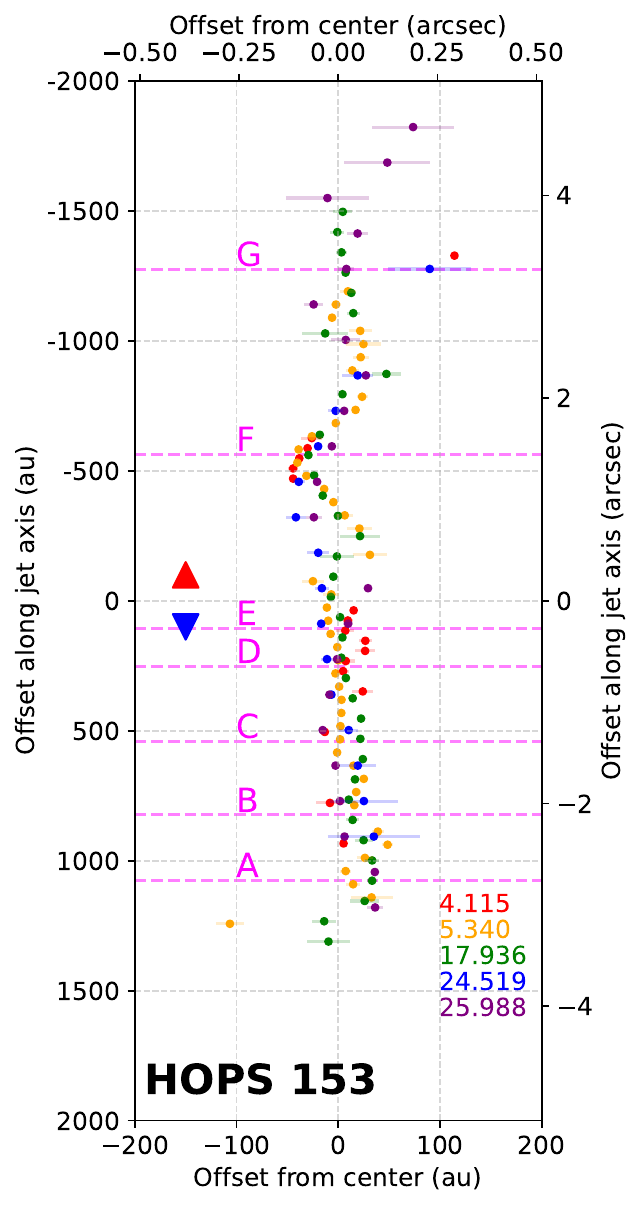}
    \includegraphics[width=.3\textwidth]{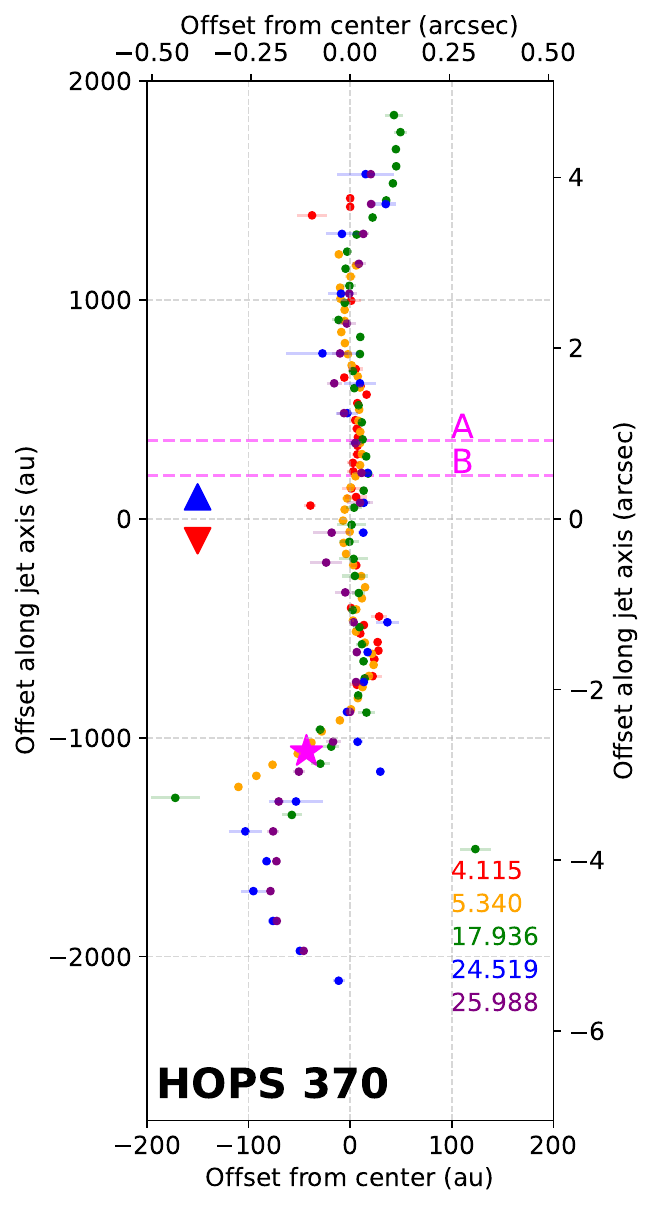}
    \caption{The deviation from the central jet axis of the centers of the Gaussian fits to the cross-section of the jets, in each of the [Fe~II] lines for the B335, HOPS~153, and HOPS~370 protostars (left to right, respectively). An arbitrary constant, individually for each line, has been added to the offsets (x-axis) to center the data points around 0~au. Blue and red triangles indicate the direction of the blue- and red-shifted lobes, respectively. In this figure and all below, positive y-axis values correspond to the blue-shifted lobe and negative values correspond to the red-shifted lobe. Horizontal magenta dashed lines show the projected distances of shocked knots listed in Table~\ref{tab:knot_pos}. In the HOPS~370 panel, a magenta star marks the position of the Class~II companion.}
    \label{fig:jet_bends}
\end{figure*}

\begin{figure*}[t]
    \centering
    \includegraphics[width=\textwidth]{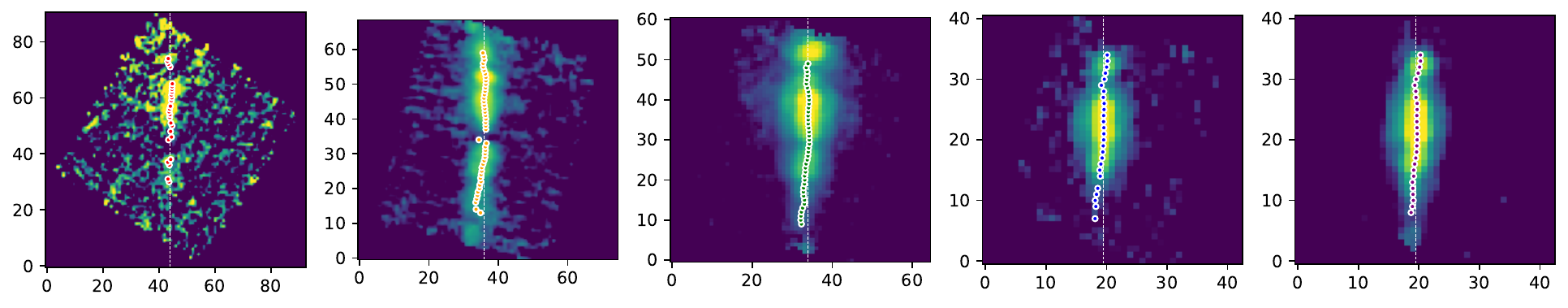}
    \includegraphics[width=\textwidth]{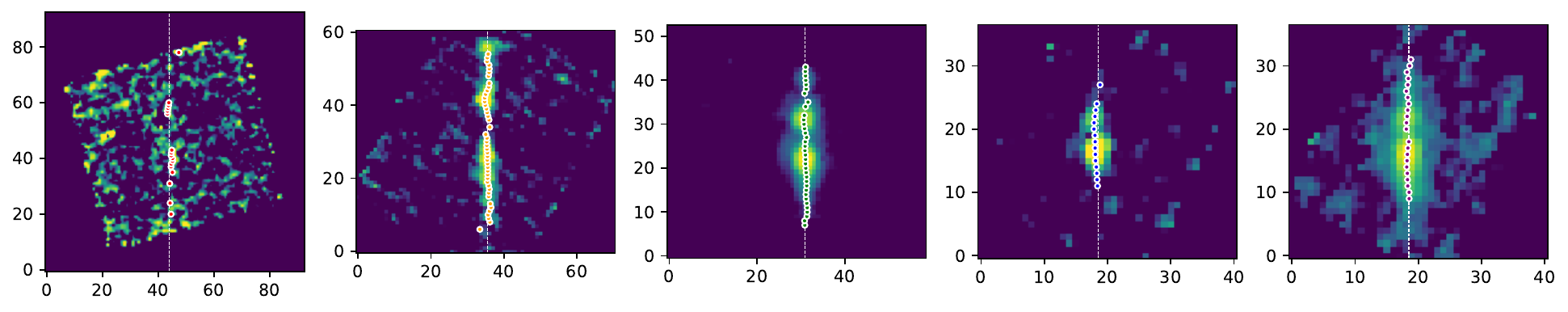}
    \includegraphics[width=\textwidth]{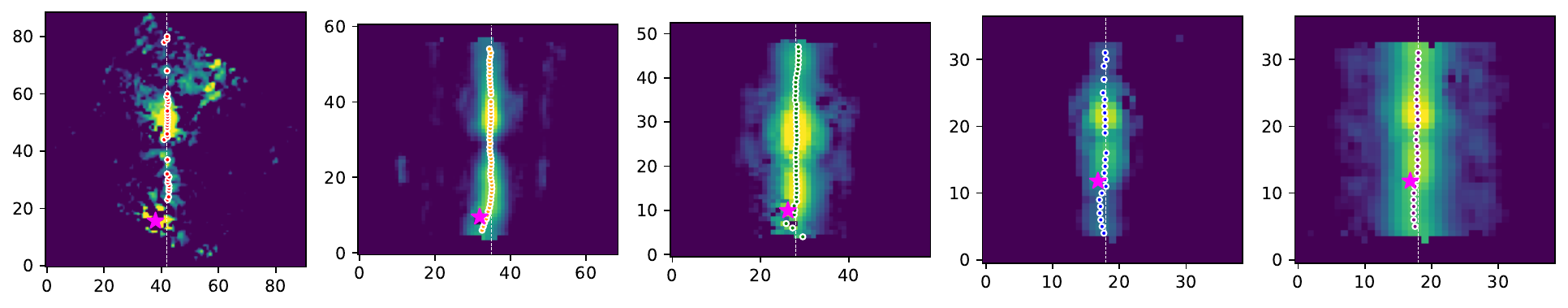}
    \caption{The deviation from the central jet axis of the centers of the Gaussian fits to the cross-section of the jets, overlaid on each of the rotated [Fe~II] line maps for the B335, HOPS~153, and HOPS~370 protostars (top to bottom, respectively). From left to right in each row are the 4.115, 5.340, 17.936, 24.519, and 25.988~$\mu$m line maps. The assumed central jet axis is shown by a vertical white dashed line.}
    \label{fig:jetbend_maps}
\end{figure*}

To map wiggles and bends in the jet and measure differences in launch angle between blue and red-shifted lobes, we first rotate each of the [Fe~II] jet maps to align the jet axis vertically using the jet position angles listed in Table~\ref{tab:sources}. We then make Gaussian fits, plus a line to account for any extended emission, to the profile of the cross-section across the jets. The fits are done along each row of pixels that extends across a cutout region around the vertically-rotated jet. In Figure~\ref{fig:jet_bends}, we plot the deviation of the Gaussian centers from the central jet axis for each of the [Fe~II] lines. The horizontal axis in Figure~\ref{fig:jet_bends} is zoomed in to emphasize the deviations from the central jet axis; Figure~\ref{fig:jetbend_maps} shows the same data points overlaid on the rotated line maps.

In HOPS~153 and HOPS~370, there is no significant deviation from 180$^{\circ}$ bipolar symmetry between the launch angle of the blue and red-shifted lobes. In contrast, B335 shows a clear deviation from 180$^{\circ}$ bipolar symmetry (Figure~\ref{fig:jet_bends}). For B335, we make linear fits to the points in both the blue-shifted and red-shifted lobes and calculate the angle between them to quantify the bend in the jet at the launch point (i.e. the protostar position); the angle between the blue- and red-shifted lobes is 168.5$^{\circ}~\pm~0.3^{\circ}$, an 11.5$^{\circ}~\pm~0.3^{\circ}$ deviation from 180$^{\circ}$ bipolar symmetry in launch angle. 

All three jets show evidence of wiggles; in the blue-shifted lobe of B335 and the red-shifted lobe of HOPS~153, the peaks of these wiggles are coincident with the position of shocked knots. In the blue-shifted lobe of the B335 jet, there is a distinct sinusoidal curve from 0 to 500~au with an amplitude ($A$) of 7~au and a wavelength ($\lambda$) of 260~au, equivalent to an angular deflection (where angle = tan$^{-1}$($\frac{4 A}{\lambda}$)) of 6.2$^{\circ}$. This angle is a "half-precession angle", analogous to a half-opening angle, from the central axis (i.e. zero deflection) to maximum deflection in one direction. Additionally, there is a $\sim$40~au deflection to the north from the central axis of the blue-shifted lobe of the B335 jet from 600 to 900~au. The red-shifted lobe of HOPS~153 shows similar wavelike wiggles from 0 to -800~au with an amplitude of 40~au and a wavelength of 650~au, equivalent to an angular deflection of 13.8$^{\circ}$. The blue-shifted lobe of HOPS~370 also has periodic wiggles from 0 to 1300 au with an amplitude of 10~au and a wavelength of 1000~au, equivalent to an angular deflection of 2.3$^{\circ}$. There is a large deflection to the west in the blue-shifted lobe of HOPS~370 from 1300 to 1800~au with an amplitude of 40~au. The red-shifted lobe of HOPS~370 shows a bend to the west from -100 to -500~au with an amplitude of 10~au, and another bend to the west from -500 to -900~au with an amplitude of 20~au. The large bend to the east ($\sim$80~au amplitude from -1000 to -2000~au) in the red-shifted lobe of HOPS~370 is coincident with a bright, more evolved companion, marked by a magenta star in Figures~\ref{fig:jet_bends} \& \ref{fig:jetbend_maps}. 

The wiggles observed are not symmetric between blue- and red-shifted lobes for the protostars in the sample. The blue-shifted lobe of B335 shows wiggles while the red-shifted lobe does not, in addition to the deviation from 180$^{\circ}$ bipolar symmetry. For HOPS~153, the red-shifted lobe shows evidence of wiggles while the blue-shifted lobe does not. HOPS~370, by contrast, shows wiggles in both the blue- and red-shifted lobes, but these are not symmetrical. The wiggles and asymmetries are discussed in Section~\ref{sec:jet_discussion}.

\subsubsection{Jet Width}\label{sec:jetwidth}

\begin{figure*}[t]
    \centering
    \includegraphics[width=0.3\textwidth]{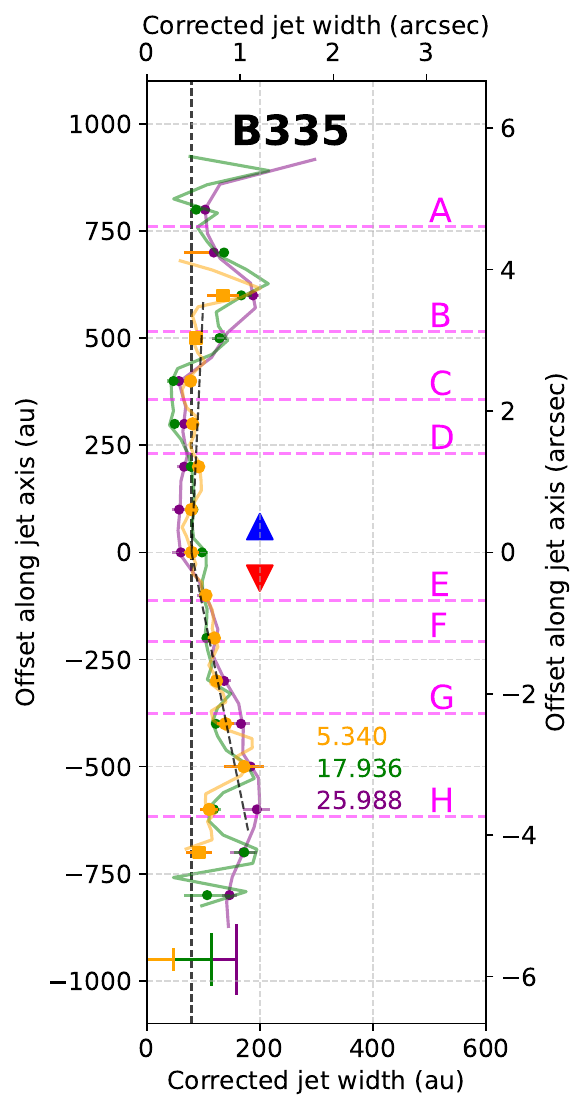}
    \includegraphics[width=0.299\textwidth]{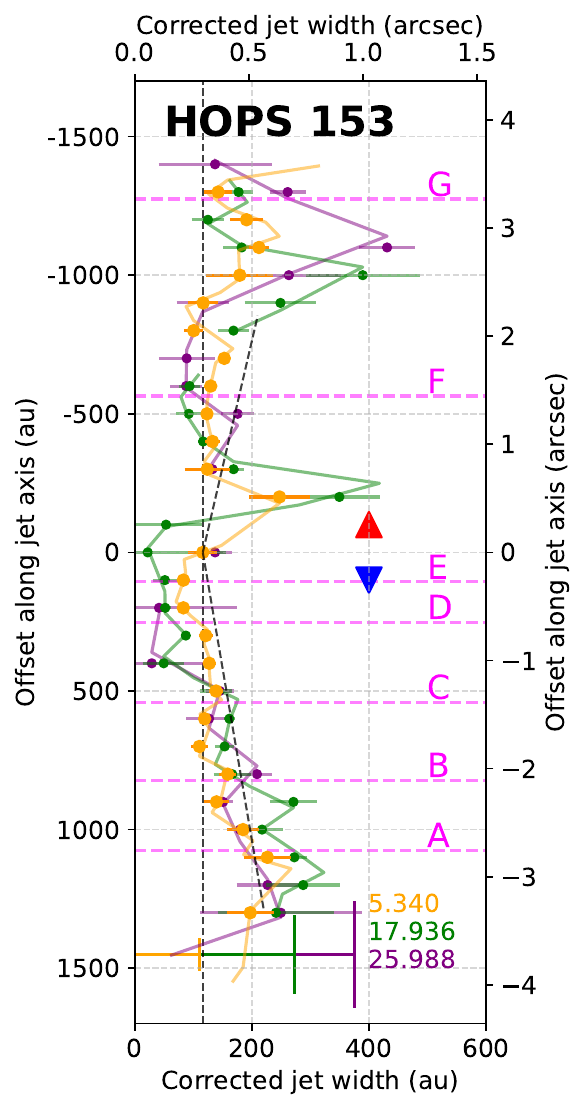}
    \includegraphics[width=0.296\textwidth]{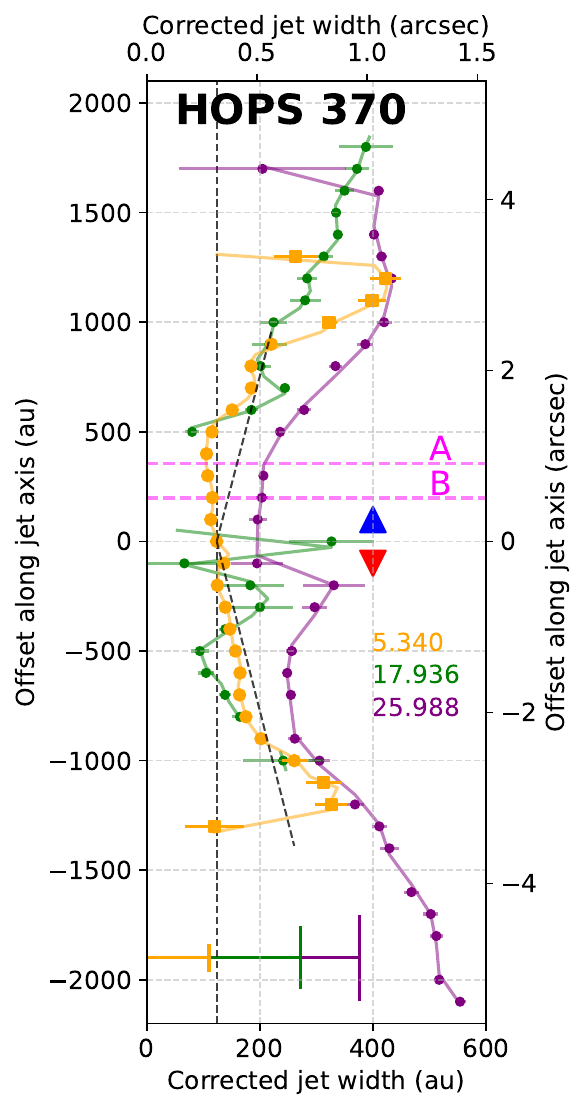}
    \caption{The widths across the vertically-aligned jet in the 5.340, 17.936, and 25.99~$\mu$m [Fe~II] lines for the B335, HOPS~153, and HOPS~370 protostars (left to right, respectively). Positive y-axis values correspond to the blue-shifted lobe and negative values correspond to the red-shifted lobe. The physical size of the MIRI MRS PSF in each wavelength is represented at the bottom of each panel by a horizontal and vertical bar. The angular resolution of the data has been subtracted in quadrature from the FWHM width of Gaussian fits across the vertically-aligned jet. Colored lines show the PSF-corrected widths, and the points show the mean widths calculated in bins of offset along the jet axis with size 100~au. Linear fits to the mean values of the 5.340~$\mu$m line are shown by black dashed lines. Mean values excluded from the linear fits of the B335 and HOPS~370 jets are shown by square points. Horizontal magenta dashed lines show the projected distances of shocked knots listed in Table~\ref{tab:knot_pos}.}
    \label{fig:deconv_jet_width}
\end{figure*}

\begin{figure*}[t]
    \centering
    \includegraphics[width=\textwidth]{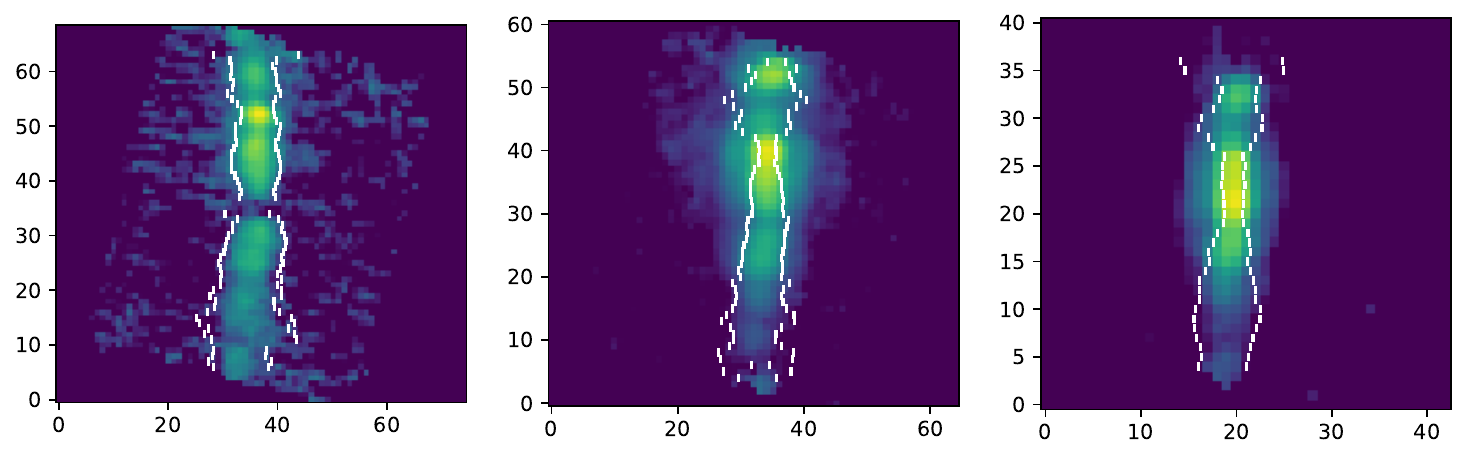}
    \includegraphics[width=\textwidth]{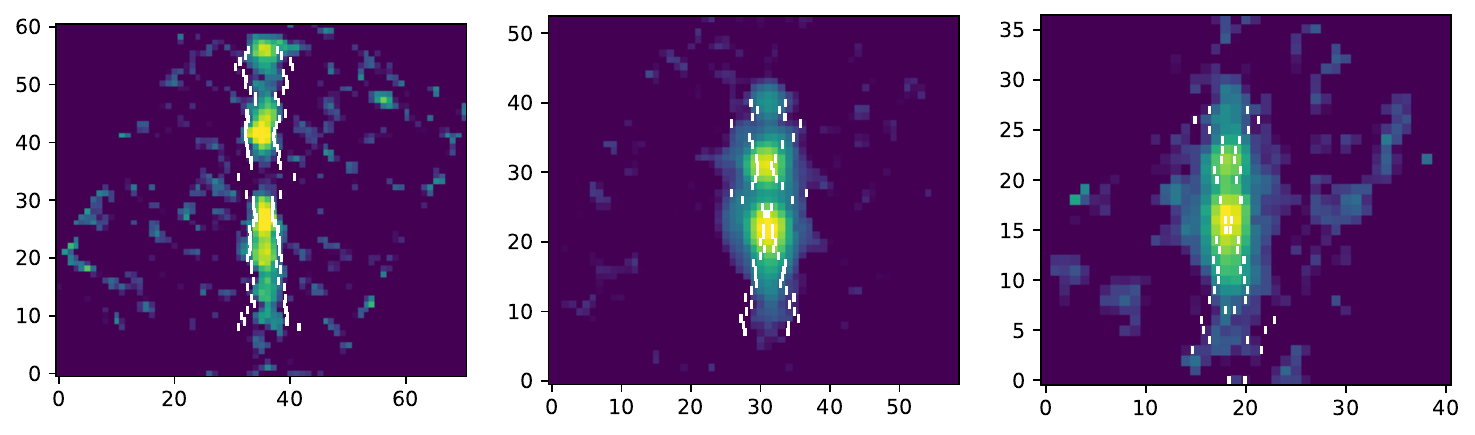}
    \includegraphics[width=\textwidth]{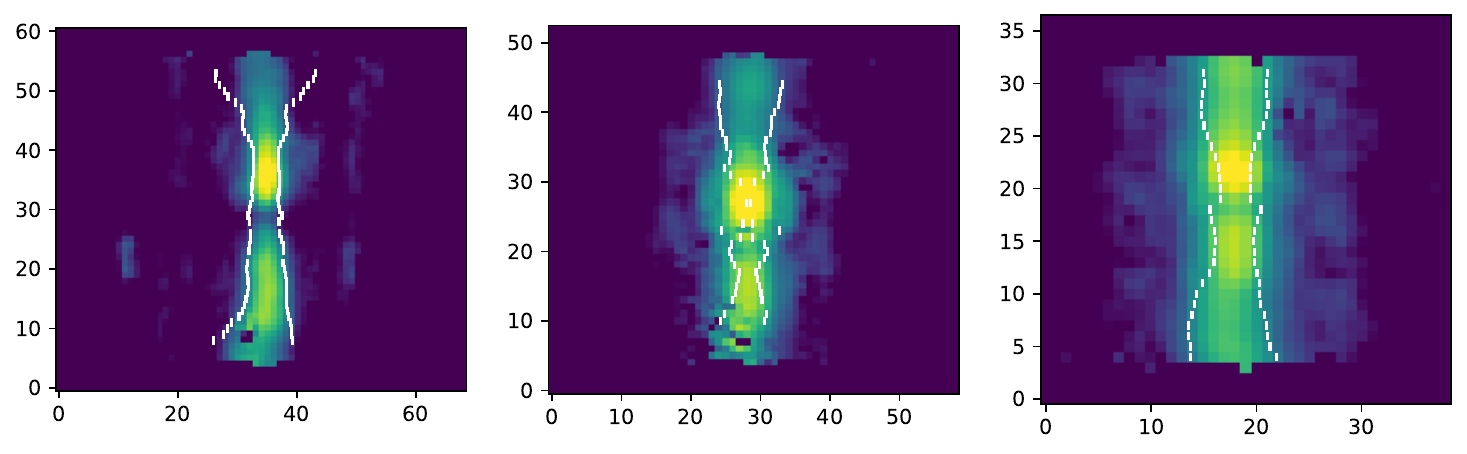}
    \caption{The widths across the vertically-aligned jet in the 5.340, 17.936, and 25.99~$\mu$m [Fe~II] lines for the B335, HOPS~153, and HOPS~370 protostars (top to bottom, respectively). From left to right in each row are the 5.340, 17.936, and 25.988~$\mu$m line maps. The fit FWHM widths are marked by white lines bounding the jets.}
    \label{fig:jetwidth_maps}
\end{figure*}

It has long been observed that protostellar jets are collimated from launch with opening angles of a few degrees, remaining collimated out to parsec scales \citep[e.g.][]{Mundt_1991, Ray_2007, Frank_2014, Pascucci_2022, Narang_2024, Caratti_2024}. 
We use the fits to each horizontal row of pixels in the vertically rotated jet images to measure the change in width along the blue and red-shifted lobes, and determine the opening angles in the inner jets. For this analysis we use only the 5.340, 17.936, and 25.988~$\mu$m [Fe~II] lines, which have stronger signal compared to the 4.115 and 24.519~$\mu$m lines. Because the widths of the jets are comparable to the angular resolution of the data in each wavelength, we subtract in quadrature the size of the MIRI PSF from the FWHM width determined from the Gaussian fits, i.e. 
\begin{equation}
\rm{FWHM}_{\rm corr.} = \sqrt{\rm{FWHM}_{\rm fit}^{2}-\rm{FWHM}_{\rm PSF}^{2}}~. 
\end{equation}
The FWHM of the MIRI PSF at each wavelength (Table~\ref{tab:resolution}) was determined from Equation 1 in \citet{Law_2023}. In Figure~\ref{fig:deconv_jet_width}, the colored lines show the PSF-corrected width of the jets for rows where the amplitude of the Gaussian fit is greater than 10 times the uncertainty in the amplitude. The colored points show the mean PSF-corrected width in bins of offset along the jet with size 100~au, and we adopt the maximum uncertainty in each bin for the errors of the average widths in each line. To examine the width of the blue- and red-shifted lobes of each jet in detail, we primarily use the mean width in the 5.340~$\mu$m line for each bin (orange points, Figure~\ref{fig:deconv_jet_width}). The 5.340~$\mu$m is the most suitable for this analysis due to the combination of strong signal and finer angular resolution, with the corrected width of the 5.340~$\mu$m jet being wider than the angular resolution at all points. However, the larger fields-of-view of the 17.936 and 25.988~$\mu$m lines provide important additional measurements at larger distances from the protostars. In Figure~\ref{fig:jetwidth_maps} we display the fit FWHM widths across each row of pixels in the rotated line maps.

We find that all the jets remain tightly collimated, and that the width of the jets vary non-monotonically over their lengths. This variation in width is not symmetric between blue- and red-shifted lobes, as described below:

\begin{itemize}

\item The width of the B335 jet at 5.340~$\mu$m is 79$\pm$13~au at the launch point, defined as the position of the protostar listed in Table~\ref{tab:sources}. The blue-shifted lobe has a relatively constant width over 0 to 500~au from launch along the jet axis, with an average width of 83~au~$\pm$~6~au. The blue-shifted lobe widens dramatically to 135$\pm$29~au at 600~au from launch, before narrowing to 59$\pm$8~au at 700~au from launch. By contrast, the red-shifted lobe of B335 widens to 174$\pm$18~au at -500~au from launch, before narrowing to 110$\pm$15~au at -600~au; the 25.988~$\mu$m line diverges from the 5.340 and 17.936~$\mu$m lines at this point.

\item The width of the HOPS~153 jet at 5.340~$\mu$m is 116$\pm$25~au at the launch point. The red-shifted lobe widens to 153$\pm$9~au from 0 to -700~au from launch, narrows to 101$\pm$17~au at -800~au, widens to 212$\pm$17~au at -1100~au, and narrows again to 142$\pm$24~au at -1300~au from launch. The 25.988~$\mu$m line diverges strongly from the 5.340 and 17.936~$\mu$m lines over -1000 to -1300~au in the red-shifted lobe. The blue-shifted lobe narrows to 83~au from 0 to 200~au, then widens to a relatively constant width over 300 to 700~au from launch along the jet axis, with an average width of 123~au~$\pm$~19~au. The blue-shifted lobe then widens to 226$\pm$41~au at 1100~au from launch. 

\item The width of the HOPS~370 jet at 5.340~$\mu$m is 123$\pm$9~au at the launch point. The blue-shifted lobe has a relatively constant width over 0 to 500~au from launch along the jet axis, with an average width of 114~au~$\pm$~6~au. The blue-shifted lobe then widens more rapidly to 422~au over 500 to 1200~au. The red-shifted lobe widens to 175$\pm$4~au over 0 to -800~au from launch, then widens more rapidly to 326$\pm$27~au over -800 to -1200~au from launch. The 25.988~$\mu$m line is wider than the 5.340 and 17.936~$\mu$m lines at all points. The changes in slope of jet width vs distance along the jet for HOPS~370 indicate that the opening angle of jets can vary over relatively short, decade timescales, and that in the recent past the jet was launched with a wider opening angle, which narrowed over time until reaching a quasi-steady state. 

\end{itemize}

To measure the opening angles of the jets, we also make a linear fit to the mean jet width observed at 5.340~$\mu$m as a function of distance from the protostar. We anchor the fits to both the blue- and red-shifted lobes at the protostar position (i.e. offset along the jet = 0~au) for all three protostars. We restrict the fit to the inner regions of the jets for B335 and HOPS~370, which show more complicated morphologies at larger distances that are not fit well by a single line. For B335, we fit the blue-shifted lobe from 0 to 500~au and the red-shifted lobe from 0 to -500~au, and for HOPS~370 we fit the blue-shifted lobe from 0 to 900~au and the red-shifted lobe from 0 to -900~au; excluded points are marked by colored squares in Figure~\ref{fig:deconv_jet_width}. We measure the opening angle of the jets by taking half of the slope for each of the fits, as described below:

\begin{itemize}

\item The blue-shifted lobe of B335 has a slope of 0.035$\pm$0.632, corresponding to an opening angle of $<$10.1$^{\circ}$. The red-shifted lobe has a slope of -0.176$\pm$0.060, corresponding to an opening angle of 5.0$^{\circ}~\pm~$0.9$^{\circ}$. 

\item The red-shifted lobe of HOPS~153 has a slope of 0.067$\pm$0.288, corresponding to an opening angle of $<6.0^{\circ}$. The blue-shifted lobe has a slope of -0.089$\pm$0.268, corresponding to an opening angle of $<6.3^{\circ}$. 

\item The blue-shifted lobe of HOPS~370 has a slope of 0.101$\pm$0.405, corresponding to an opening angle of $<8.7^{\circ}$. The red-shifted lobe has a slope of -0.072$\pm$0.069, corresponding to an opening angle of 2.1$^{\circ}~\pm~$1.0$^{\circ}$. 

\end{itemize}

In most cases , we are only able to provide upper limits to the opening angles as the the uncertainties in the opening angles are large relative to their values. Only the opening angles of the red-shifted lobes of B335 and HOPS~370 can be accurately constrained. The large errors on the opening angles highlight the fact that the jets have significant sub-structure, and fitting opening angles from a single line even to the inner parts of the jets are a very crude estimate. While these fits demonstrate a general trend for the jets to widen with distance from the launching point, this widening is complex and variable; we discuss this further in Section~\ref{sec:jet_discussion}.

\subsection{Jet Kinematics}\label{sec:kinematics}

We explore the $V_{rad}$ structure of the jets through position-velocity (P-V) diagrams, comparing the velocity difference of the data relative to the lab wavelength of each [Fe~II] line versus position along the jet. The overall wavelength calibration of MIRI MRS is accurate to within 5-10~km~s$^{-1}$, depending on the sub-channel \citep{Argyriou_2023, Pontoppidan_2024b,Banzatti_2025}. 
For a single line, however, much smaller velocity differences can be measured by searching for shifts in the velocity \citep[e.g.][]{Narang_2024}. This enables an estimation of the jet $V_{rad}$ and variations in the velocity through the relative difference between the blue-shifted lobe and the red-shifted lobe, assuming symmetric velocities in the blue- and red-shifted lobes. 

In principle, we should not assume that the blue- and red-shifted lobes have symmetrical velocities, especially in light of the various asymmetries apparent in their morphologies. However, when we neglect this assumption and instead measure the velocities relative to the Local Standard of Rest or systemic velocities, we obtain non-physical results (Appendix~\ref{sec:vel_offsets}). These results suggest there is an additional systematic uncertainty in the velocities, beyond the calibration uncertainties listed in the JWST documentation. For this reason, we adopt the analysis technique of \citet{Narang_2024}, assuming symmetrical velocities between blue- and red-shifted lobes. 

\begin{figure*}[t]
    \centering    
    \includegraphics[width=0.25\textwidth]{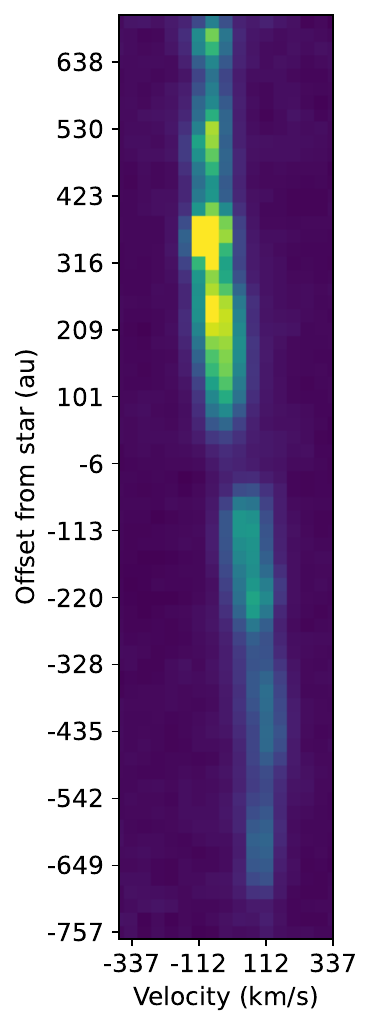}
    \includegraphics[width=0.248\textwidth]{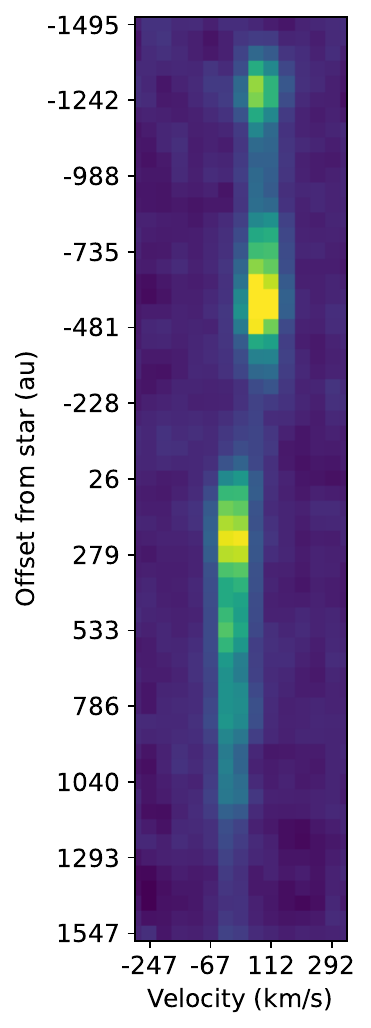}
    \includegraphics[width=0.2663\textwidth]{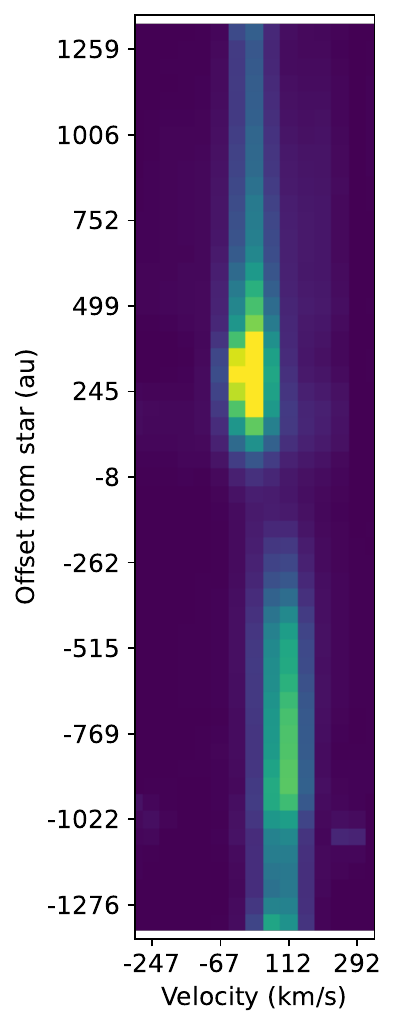}
    \caption{Position-velocity diagrams for the 5.340 $\mu$m [Fe~II] line for the B335, HOPS~153, and HOPS~370 protostars (left to right, respectively). The vertical axis is the projected distance to the protostar along the jet axis (Table~\ref{tab:sources}). Positive offsets correspond to blue-shifted lobes and negative offsets correspond to red-shifted lobes.}
    \label{fig:5.34_pv}
\end{figure*}

As for the jet widths in Section~\ref{sec:jetwidth}, we use the 5.340, 17.936, and 25.988~$\mu$m [Fe~II] lines. We start with the continuum-subtracted data cubes, rotated vertically as for the summed maps in the preceding sections, and collapse the 3D cubes to 2D by taking the mean along the horizontal axis (i.e. collapsing across the width of the jet to create 2D maps in position-velocity space, with position being along the jet axis). We then calculate the velocity of each pixel in the spectral axis by taking the difference from the lab wavelengths of the [Fe~II] lines. The P-V diagrams for the 5.340 $\mu$m line are displayed in Figure~\ref{fig:5.34_pv}; the 5.340 $\mu$m line has both higher angular and spectral resolution, although with limited FOV. 

\begin{figure*}[t]
    \centering
    \includegraphics[height=10cm]{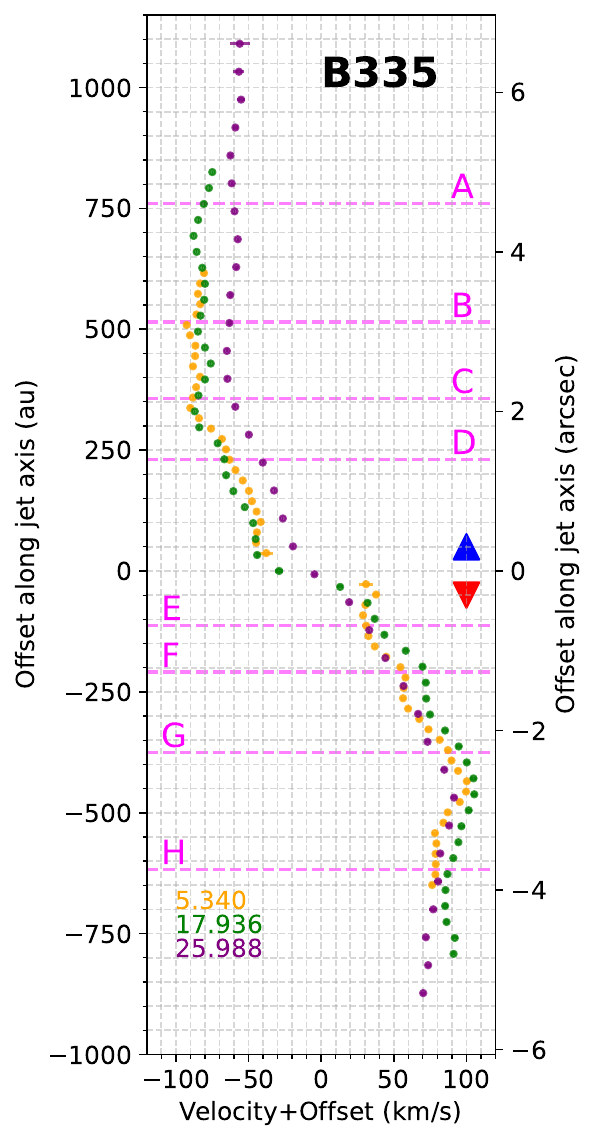}
    \includegraphics[height=10cm]{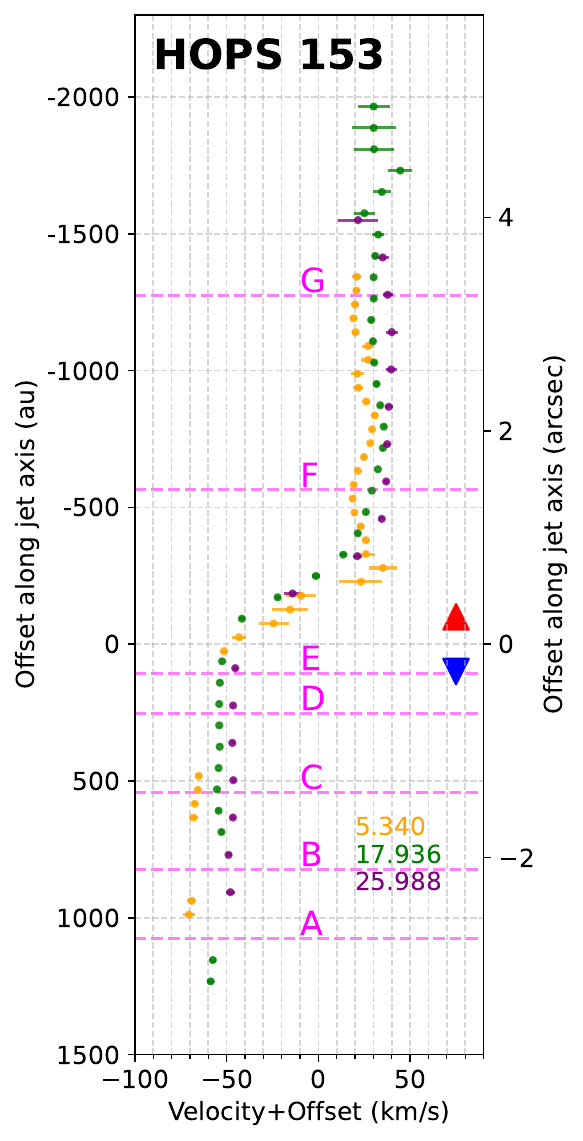}
    \includegraphics[height=10cm]{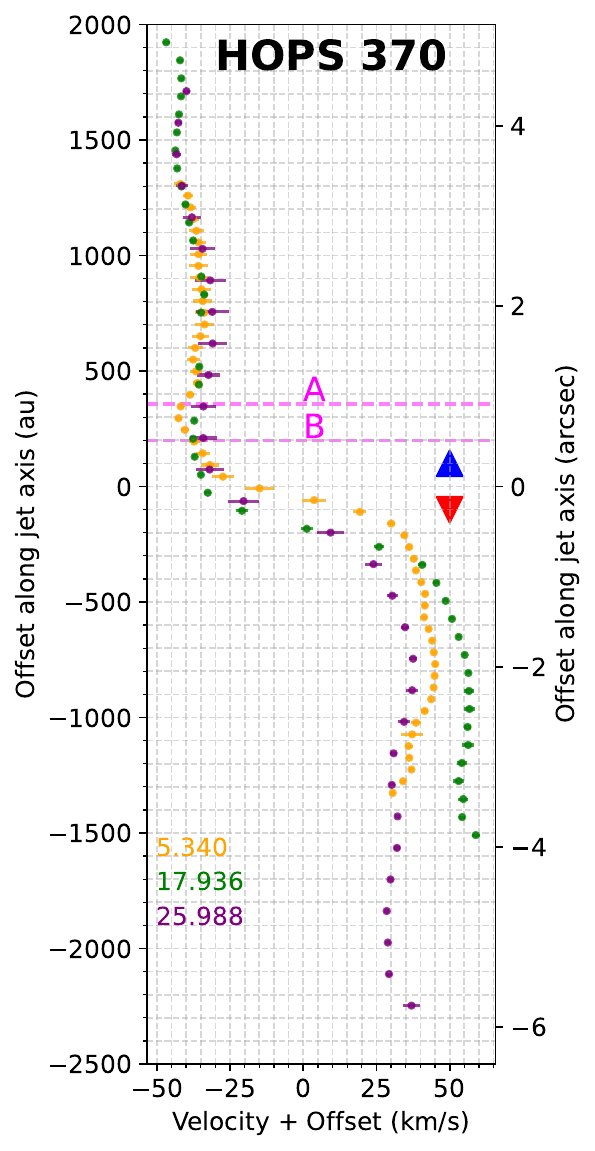}
    \caption{The centers of the Gaussian fits to the P-V diagrams for each of the [Fe~II] lines accessible by MIRI for the B335, HOPS~153, and HOPS~370 protostars (left to right, respectively). An offset has been added to center the velocity around 0~km~s$^{-1}$; half the difference between the blue and red-shifted lobes gives the jet velocity, assuming symmetric velocities. Positive y-axis values correspond to the blue-shifted lobes, and negative y-axis values correspond to the red-shifted lobes. Horizontal magenta dashed lines show the projected distances of shocked knots listed in Table~\ref{tab:knot_pos}.}
    \label{fig:pv_fits}
\end{figure*}

We fit a Gaussian profile plus a linear baseline across the spectral axis for each row of pixels along the vertical axis of the P-V diagram. In Figure~\ref{fig:pv_fits}, we plot the centers of the Gaussian fits to the velocity axis versus distance along the blue and red-shifted lobes for the [Fe~II] lines accessible by MIRI. Because the absolute wavelength calibration is uncertain \citep{Argyriou_2023,Pontoppidan_2024b,Banzatti_2025}, we offset the velocities so that the mean velocity of all the measurements is set to 0~km~s$^{-1}$.

In Table~\ref{tab:jetvelocity}, we list the $V_{rad}$ calculated from half the difference between the maximum and minimum velocity in each of the [Fe~II] lines accessible by MIRI, as well as an average of all the lines, uncorrected for disk inclination. This assumes the blue and red-shifted lobes have symmetric velocities, as was done for IRAS~16253 \citep{Narang_2024}. The protostar position of HOPS~153, determined from ALMA, is offset from the position where $V_{rad}$ = 0~km~s$^{-1}$; this is due to the high extinction to the northwest of the ALMA position that obscures the base of the red-shifted lobe, biasing the average velocity toward the ALMA position. For each line in the table, we adopt the maximum uncertainty in the fit velocities for that line, and for the average velocity we adopt the sum of the maximum uncertainties in each line in quadrature.

\subsubsection{Total 3D Jet Velocities}\label{sec:inclination}

In the final two columns of Table~\ref{tab:jetvelocity}, we list the jet tangential velocity along the length of the jet in the plane of the sky ($V_{tan}$) and total velocity ($V_{tot}$, i.e. the inclination-corrected deprojected velocity), found from the average $V_{rad}$ corrected for the disk inclination angles listed in Table~\ref{tab:sources}, where $V_{tot} = V_{rad}/\textrm{cosine(inclination})$. However, correcting the average $V_{rad}$ for the inclination angle of B335 provides an unrealistically high $V_{tan}$ of 793~$\pm$~77~km~s$^{-1}$ and $V_{tot}$ of 799~$\pm$~78~km~s$^{-1}$. With an estimated stellar mass of 0.25~M$_{\odot}$ \citep{Evans_2023}, an 800~km~s$^{-1}$ total jet velocity would require a vanishingly small launch radius of 0.15~R$_{\odot}$ under the assumption that the launch velocity is equal to the Keplerian velocity. 
We discuss this discrepancy in the jet velocity and inclination of B335 in detail below.

\begin{table*}[t]
\caption{Radial, tangential, and total velocities of the [Fe~II] jets observed with MIRI MRS}
\centering
\begin{tabular}{lcccccc}
    {} & {$V_{rad,~5.340}$} & {$V_{rad,~17.936}$} & {$V_{rad,~25.988}$} & {$<$$V_{rad}>$} & {$V_{tan}$} & {$V_{tot}$}\\
    {Source} & {km~s$^{-1}$} & {km~s$^{-1}$} & {km~s$^{-1}$} & {km~s$^{-1}$} & {km~s$^{-1}$} & {km~s$^{-1}$} \\ \hline
    B335 & 96~$\pm$~6 & 97~$\pm$~3 & 78~$\pm$~7 & 90~$\pm$~9 & 140~$\pm$~28 & 166~$\pm$~16 \\
    HOPS 153 & 53~$\pm$~12 & 52~$\pm$~12 & 45~$\pm$~11 & 50~$\pm$~20 & 179~$\pm$~72 & 186~$\pm$~75 \\
    HOPS 370 & 44~$\pm$~5 & 53~$\pm$~2 & 40~$\pm$~6 & 46~$\pm$~8 & 142~$\pm$~24 & 150~$\pm$~26 \\
\end{tabular}
\label{tab:jetvelocity}
\end{table*}

The inclination of the B335 system estimated by fitting radiative transfer models to the SED is 87$^{\circ}$ \citep[Section~\ref{sec:b335_inc} of this work,][]{Stutz_2008}. The inclination estimated from the large-scale outflow observed in CO (J=1-0) at 115 GHz is 80$^{\circ}$ \citep{Hirano_1988}. We adopt the average of these two values, which coincidentally best reproduces the JWST spectrum in radiative transfer modeling. The unrealistically high velocity of the B335 jet after correcting for an 83.5$^{\circ}$ inclination motivates a deeper exploration of the inclination of the jet itself, which may differ from that of the outflow cavity \citep[e.g.][]{Hull_2016}. 

\begin{figure}[ht]
    \centering
    \includegraphics[scale=0.65]{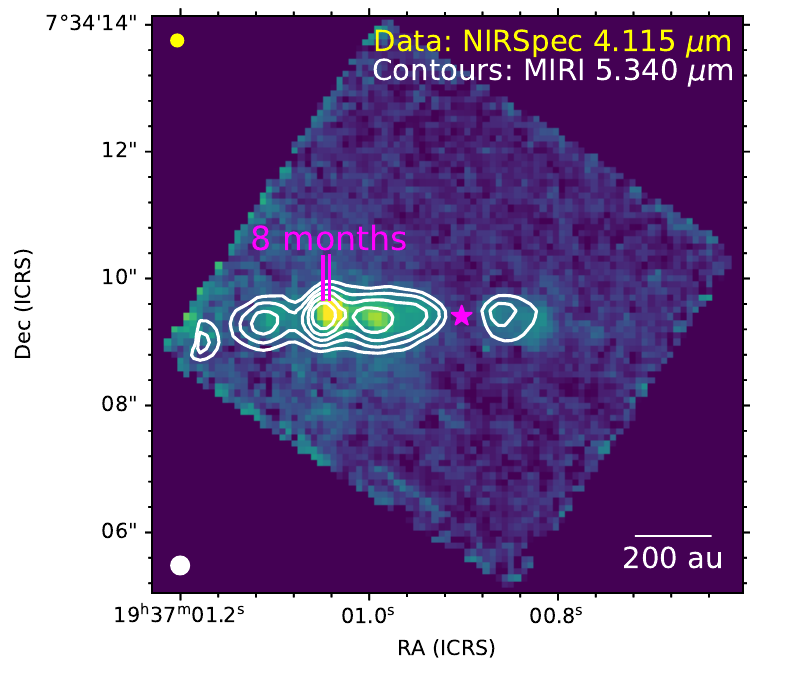}
    \caption{A demonstration of the method to determine the tangential velocity of the B335 jet, from the position of a shocked knot (knot B335-C) in the blue-shifted lobe. The data show the 4.115~$\mu$m [Fe II] line from NIRSpec, while the white contours show the 5.340~$\mu$m [Fe II] line from MIRI. The tangential velocity was determined from the positions of the brightest knot. The difference between the fits to the Gaussian centers of the shocked knot indicated by the pink lines, divided by 8 months, provides the proper motion. The star position is marked by a magenta star, the NIRSpec PSF is shown in yellow at the top left, and the MIRI PSF is shown in white at the bottom left. A 200~au scalebar is shown in white at the bottom right.}
    \label{fig:knotpm}
\end{figure}

Such an exploration is possible for B335 due to its relative proximity to the Sun (d = 165~pc) and an 8 month gap between the NIRSpec and MIRI JWST observations for this source, enabling a measure of the tangential motion of shocked knots in the jet. We determine the center of the brightest shocked [Fe~II] knot (knot B335-C) in both NIRSpec (4.115~$\mu$m) and MIRI (5.340~$\mu$m) by fitting 2D Gaussian profiles to the knot in each line. The distance between the center of the knot in NIRSpec and MIRI, divided by 8 months, provides an estimate of the velocity of the knot in the plane of the sky (Figure~\ref{fig:knotpm}, Table~\ref{tab:knotpm}). An important caveat to this approach is the fact that the proper motion of a shocked knot may not be equal to the bulk velocity of the gas flowing through the knot, such as in the case of a recollimation shock in which the knot remains stationary as gas flows through \citep[e.g.][]{Eisloffel_1992}. Additionally, there is a possibility that the different lines are tracing different physical regions of the shock \citep[e.g.][]{Dutta_2025}; however, as the lines are from the same species with similar excitation energies, any such difference should not be significant relative to the uncertainties in the fitting.

We measure a proper motion of 0\farcs177~$\pm$~0\farcs035 per year, which translates to a tangential velocity ($V_{tan}$) of 140~$\pm$~28~km~s$^{-1}$ \citep[consistent with the 129~km~s$^{-1}$ tangential velocity of][using JWST GTO NIRCam data]{Hodapp_2024}. This, combined with the $V_{rad}$ of 90~$\pm$~9~km~s$^{-1}$, implies a more realistic total jet velocity of 166~$\pm$~16~km~s$^{-1}$ and a jet inclination angle of 57$^{\circ}$~$\pm$~2$^{\circ}$. In an investigation of the high-velocity CO outflow observed with ALMA, \citet{Kim_2024} determined that the high-velocity CO was likely ejected at a velocity of 156~km~s$^{-1}$, very close to our derived total velocity for the [Fe~II] jet; however, this velocity was calculated using the 80$^{\circ}$ inclination from \citet{Hirano_1988}. This knot has an estimated launch date between MJD 56372 and 57320. In more recent ALMA observations of the CO knot, the proper motion and radial velocityresults in an inclination angle of 48$^{\circ}$, resulting in a launch date of $\sim$MJD 57750 (Kim Chul-Hwan 2025, private communication). The apparent misalignment between the jet, CO outflow components, and cavity of B335 could be due to motion of the jet along the line of sight, as discussed in the following section.

\begin{table*}[t]
\caption{Proper motion from 2D Gaussian fits to a shocked [Fe~II] knot B335-C seen in 4.115~$\mu$m (NIRSpec) and 5.340~$\mu$m (MIRI). 
}
\centering
\begin{tabular}{ccc}
    NIRSpec (RA, Dec) & ($294.254330^{\circ}, 7.569295^{\circ}$) & ($\pm$0\farcs01, $\pm$0\farcs01) \\ \hline
    MIRI (RA, Dec) & ($294.254359^{\circ}, 7.569282^{\circ}$) & ($\pm$0\farcs02, $\pm$0\farcs02) \\ \hline
    $\Delta_{Coordinate}$ ($\Delta_{RA}$, $\Delta_{Dec}$) & (0\farcs106, -0\farcs050) & ($\pm$0\farcs023, $\pm$0\farcs023) \\ \hline
    $\Delta_{total}~('')$ & 0.117 & $\pm$0.023 \\ \hline
    Proper Motion ($''~yr^{-1}$) & 0.177 & $\pm$0.035 \\ \hline
    Proper Motion ($au~yr^{-1}$) & 29.3 & $\pm$5.8 \\ \hline
    Proper Motion ($km~s^{-1}$) & 140 & $\pm$28 \\ \hline
\end{tabular}
\label{tab:knotpm}
\end{table*}

\section{Discussion}\label{sec:jet_discussion}

In this section, we discuss the implications of the results detailed in Section~\ref{sec:iron_jet_results}. The morphologies and kinematics of the jets within 1000~au of their launch provide critical context for the physical conditions of their launch. The complexity of the data is a testament to the superlative capabilities of JWST in the observation of young, deeply embedded protostars undergoing accretion and outflow. Although one of the goals of the IPA program was to trace variations in outflows with protostellar mass, we do not discern any obvious trend in jet morphology or kinematics with mass in the three objects analyzed. However, this is not surprising given the small number of objects in the sample and the very stochastic nature of protostellar accretion and outflow, which are highly sensitive to factors such as environment and disk properties.



\subsection{Jet Widths and Opening Angles}\label{width_discussion}

With the MIRI IFU data we are able to resolve the widths of the jet to within 100~au of the launching point, and the change in width as they propagate away from the source. This enables a measurement of the opening angles of the jets, and the implied tangential velocity across the width of the jet. The widths are $\sim$100~au at their launching point, widening non-monotonically to 200-400~au at 1000~au from the launch point. The widths of the jets near the launch point are independent of the sizes of the protostellar disks.

The jet width as a function of distance shows significant variations that are not described by the monotonic growth as found in other data (Section~\ref{sec:jetwidth}, Figure~\ref{fig:deconv_jet_width}). The horizontal dashed lines in Figure~\ref{fig:deconv_jet_width} mark the projected distance to shocked knots identified in Table~\ref{tab:knot_pos} from the protostars. The shocked knots in the jet of B335 at -617, +231, +357, and +761~au (knots B335-H, D, C, and A) are all coincident with regions where the jet becomes narrower, while the region where the jet becomes dramatically wider at +600~au is coincident with the space between two knots. The blue-shifted lobe of HOPS~153 also narrows near the position of the shocked knot at +106~au (knot HOPS~153-E), and the jet of HOPS~370 is tightly collimated at the position of the two shocked knots in the blue-shifted lobe, beyond which the jet becomes wider. This implies that the profiles of the jet are more sharply peaked at the knots than in the spaces between the knots. 

Additionally, the 25.988~$\mu$m [Fe~II] line deviates from the values of the 5.340 and 17.936~$\mu$m lines at the position of the shocked knot in the red-shifted lobe of B335 at -617~au (knot B335-H) and in the red-shifted lobe of HOPS~153 from -1000 to 1400~au. The 25.988~$\mu$m line is wider than the 5.340 and 17.936~$\mu$m lines at all points in the case of HOPS~370 (Figure~\ref{fig:deconv_jet_width}). We suggest that this is due to the lower angular resolution data at this wavelength being sensitive to a more extended, less collimated shell of a stratified jet as observed for the Class~0 protostar HH~211\citep{Caratti_2024}.

The opening angle of the jet from IRAS~16253 was found to be consistent with ballistic expansion (i.e. no additional confinement mechanisms beyond thermal motions) by comparing the tangential jet velocity with the sound speed, assuming a shock temperature of 10,000~K \citep[equal to a sound speed of 10~km~s$^{-1}$,][]{Narang_2024}. We adopt a different approach, calculating the inferred transverse velocity across the width of the jet (i.e. rate of expansion of the jet). The transverse velocity across the jet from the central jet axis to the edge is determined from the projected $V_{tan}$ along the length of the jet axis ($V_{tan \parallel}$, Table~\ref{tab:jetvelocity}), second column from right) and the opening angle ($\theta_{opening}$) found in Section~\ref{sec:jetwidth}. The transverse velocity across the jet, $V_{trans,\perp}$, is calculated as $V_{trans,\perp} = V_{tan,\parallel} \times \textrm{tan}(\theta_{opening})$. Because of the very large uncertainties in the opening angles, in some cases we can only provide upper limits to the transverse velocities. We find that the blue-shifted lobe of B335 has a transverse velocity across the jet of $<25.1~km~s^{-1}$, and the red-shifted lobe has a transverse velocity of 12.4~$\pm$~3.2~km~s$^{-1}$. The red-shifted lobe of HOPS~153 has a transverse velocity across the jet of $<18.9~km~s^{-1}$, and the blue-shifted lobe has a transverse velocity of $<20.0~km~s^{-1}$. The blue-shifted lobe of HOPS~370 has a transverse velocity across the jet of $<21.3~km~s^{-1}$, and the red-shifted lobe has a transverse velocity of 5.1~$\pm$~2.7~km~s$^{-1}$. These transverse velocities translate to expanding sound speeds for an ideal gas with temperatures ranging from 500~K at the lowest transverse velocity to 13000~K at the highest (for the red-shifted lobe of B335). Given the extremely high uncertainties, these measurements are only a rough diagnostic of the expansion of the jets, and more detailed comparisons to models of the physical conditions in the jets are required for further analysis. We consider, and rule out as unlikely, the possibility of external collimation of the jets due to the infalling envelope \citep[e.g.][]{Wilkin_1994, Carrasco_2021}, as the jets are propagating into a relatively low-density cavity carved out of the envelope.

\subsection{Wiggles and Bends in Protostellar Jets}\label{sec:wiggle_discussion}

Asymmetric wiggles in the blue- or red-shifted jets in the plane of the sky, tracing angular deflections of 1$^{\circ}$ - 14$^{\circ}$ in the jet launching vector, are observed for B335, HOPS~153, and HOPS~370, as detailed in Section~\ref{sec:iron_jet_results}. Such wiggles are commonly observed in precessing jets at many different size scales \citep[e.g.][]{Eisloffel_1997, Louvet_2018, Paron_2022, 2023ApJ...948...39R, Erkal_2021, Bajaj_2025}, often linked to precession of the disk. One of the most commonly invoked causes of precession from protostellar jets is through tidal interaction of the Keplerian disk with a binary companion \citep[e.g.][]{Fendt_1998, Terquem_1999, Bate_2000, Masciadri_2001, Sheikhnezami_2015, Louvet_2018, Erkal_2021, Paron_2022, 2023ApJ...948...39R, Bajaj_2025}. If the binary is inclined with respect to the disk plane, the gravitational torque leads to a differential precession within the disk, resulting in a misalignment between different regions of the disk, called a warp \citep[e.g.][]{Facchini_2013, Zanazzi_2018, Deng_2022}. There is currently no evidence of binary companions for the three protostars studied in this work, and such precession is usually symmetric between blue- and red-shifted lobes. Nonetheless, the presence of a close undetected binary companion or massive protoplanet remains a possibility. 

A possible cause of wiggles in protostellar jets is a warping of the disk, altering the angular momentum vector and thus the jet launch vector. Models suggest multiple physical processes that can induce protostellar disk warping, besides interaction with a binary companion. One such process is anisotropic infall, where a large amount of material falls asymmetrically onto the disk, warping the disk and altering the angular momentum vector \citep[e.g.][]{Shepherd_2000, Hirano_2019, Kuffmeier_2021, Kuffmeier_2024}. Such asymmetric infall from streams of material onto protostellar disks from the surrounding envelope are becoming a commonly observed phenomenon, especially for deeply-embedded protostars \citep[e.g.][]{Ginski_2021, ValidiviaMena_2024, Speedie_2025}. 
Another possible driver of warps in protostellar disks can be interactions of the disk with a stellar magnetic field and/or the larger-scale magnetic field threaded through the disk \citep[e.g.][]{Lai_2003, Fendt_2009, Fendt_2022}. In typical circumstellar disk conditions, warps travel in a wave through the disk \citep{Lubow_2000}, which can also lead to a disk precession \citep{Kimmig_2024}, changing the orientation of the orbital planes within the disk. However, in such cases the models predict symmetric wiggling between blue- and red-shifted lobes, contrary to our results. We investigate the possibility of warping of the inner disks relative to the outer disks below, but do not claim that such warping is responsible for the observed asymmetric wiggles in the jets.

As the collimated jet is launched from the innermost regions of the disk, this warping could result in a misalignment of the jet axis from 90$^{\circ}$ to the dust disk major axis as observed at (sub-)mm wavelengths. To investigate this, we compare the position angles (P.A.) of the major axes of the dust disks detected in the (sub-)mm listed in \citet{Rubinstein_2024} to the P.A. of the collimated iron jets. For this analysis, we utilize the jet angles listed in \citet{Federman_2024}, which were determined manually by eye; we therefore assume a conservative error in the jet P.A. of 25\%. B335 has a jet P.A. of 90$^{\circ}$~$\pm$~23$^{\circ}$ and a dust disk major axis P.A. of 5$^{\circ}$ \citep{Bjerkeli_2019}, with an 85$^{\circ}$ alignment between disk and jet axes. HOPS~153 has a jet P.A. of -55$^{\circ}$~$\pm$~14$^{\circ}$ and a dust disk major axis P.A. of 33$^{\circ}$ \citep{Tobin_2020A}, with an 88$^{\circ}$ alignment between disk and jet axes. HOPS~370 has a jet P.A. of 6$^{\circ}$~$\pm$~2$^{\circ}$ (consistent with the inclination of the cm jet observed with VLA, Appendix~\ref{sec:hops370_vla}) and a dust disk major axis P.A. of 110$^{\circ}$ \citep{Tobin_2020A}, with a 104$^{\circ}$ alignment between disk and jet axes. For B335 and HOPS~153, the alignment between disk and jet axes in the plane of the sky are roughly consistent with 90$^{\circ}$. By contrast, HOPS~370 is 14$^{\circ}$ off from perpendicular, as seen in Figure~\ref{fig:hops370_jetmaps}. This misalignment supports the idea that there is a warping of the inner disk relative to the outer disk of HOPS~370.  

In addition to the smaller-scale wiggles, the jet of B335 shows a larger bend in the blue-shifted lobe of the B335 jet with an amplitude of $\sim$40~au. Larger-scale bends in protostellar jets have long been observed; proposed causes include interaction with a binary companion, dynamical pressure as a protostar moves through an external medium, or deflection due to Lorentz forces from a large-scale magnetic field \citep[e.g.][]{Fendt_1998}. For externally irradiated jets, other possible causes include photoablation and ionization of a dense neutral jet, radiation pressure on dust entrained in the jet, or deflection of the jet via a side-wind from an external source \citep[e.g.][]{Bally_2001, Bally_2006, Bally_2012, Masciadri_2001}. Finally, it is possible for a protostellar jet to be deflected after collision with the larger environment \citep[e.g.][]{Choi_2005, Osorio_2017}. 

For B335, which is relatively isolated and deeply embedded, the red-shifted lobe is wider than the blue-shifted lobe and gets wider with distance, with an opening angle of 5$^{\circ}$ compared to $<$1$^{\circ}$ for the blue-shifted lobe. This asymmetry in jet width, combined with the deviation of the jet from 180$^{\circ}$ bipolar symmetry, could be due to the effects of the dense protostellar disk on the stellar and/or disk magnetic field structure \citep[e.g.][]{Lovelace_2010, Dyda_2015, Fendt_2013, Fendt_2022}. The large 40~au bend in the blue-shifted lobe of the B335 jet from 600 to 900~au, by contrast, could be the result of past variations in the jet launch angle. The more massive and luminous HOPS~370, residing in a much more crowded region of star formation on the eastern edge of a dust clump observed with ALMA \citep{Osorio_2017, Tobin_2020A, Federman_2023}, has a large bend in the red-shifted lobe with an amplitude of 40~au from -500 to -1000 au. \citet{Osorio_2017} found evidence that the southern jet of HOPS~370 collides with a dense clump at a distance of 8000-10000~au to the southwest of the protostar, and evidence of precession or deflection of the northern jet at similarly large distances. This suggests that some of the structure observed in the jet of HOPS~370 could be due to interaction with the surrounding dusty molecular cloud.

\subsection{Asymmetries in Jet Morphology}\label{sec:asymmetries}

The results of Section~\ref{sec:iron_jet_results} demonstrate asymmetries between the blue-shifted and red-shifted lobes of all three jets. This asymmetry is apparent in the number and spatial distribution of shocked knots in the blue- versus red-shifted lobes. If knots are ejected symmetrically, they should have equal numbers of knots equidistant along the lobes. Although the number of shocked knots identified in both lobes of the B335 jet are equal, none of the knots have comparable distances within their uncertainties that would indicate symmetric launching. The blue-shifted lobe of HOPS~153 has four knots identified compared to two in the red-shifted lobe, and only the knots at +541~au and -564~au knots HOPS~153-F and HOPS~153-C) have comparable distances within their uncertainties. The blue-shifted lobe of HOPS~370 has two knots, with a smooth distribution for the rest of the extent of the blue-shifted lobe, and a smooth distribution with no identified knots in the red-shifted lobe. 
Asymmetries were also noted in the observed wiggles and widths of the jets between blue- and red-shifted lobes. The blue-shifted lobes of B335 and HOPS~370, and the red-shifted lobe of HOPS~153, display roughly sinusoidal wiggles while their counterparts do not. The opening angles of the blue-shifted and red-shifted lobes are comparable for HOPS~153 and HOPS~370, but for B335 the red-shifted lobe has an opening angle $>$5 times that of the blue-shifted lobe. Additionally, there is an 11.5$^{\circ}$ deviation from 180$^{\circ}$ bipolar symmetry in the launch angle between the blue- and red-shifted lobes of the B335 jet. 

Such asymmetries between blue-shifted and red-shifted lobes have been observed in visible and infrared wavelengths in many protostellar jets \citep[e.g.][]{Xie_2021, Asssani_2024, vanDishoeck_2025}. We discussed above in Section~\ref{sec:wiggle_discussion} the possibility of the protostellar magnetic field (which can be more complicated than a simple dipole), and/or larger-scale magnetic field threaded through the disk, being responsible for warps in the disk which drive precession of the jet launching axis. Magnetohydrodynamic simulations of protostellar systems including these interactions between magnetic fields and disks can produce asymmetric precession and opening angles between red and blue-shifted lobes \citep[e.g.][]{Fendt_2013, Mori_2025}. The fact that the launching of protostellar jets is mediated by magnetic fields in the inner disk strongly suggests that magnetic fields play a significant role in the observed asymmetries in protostellar jets. However, the immediate surrounding environment can also play a role, as the blue- and red-shifted lobes may propagate into different density media \citep[as in the case of HH~46/47][]{Birney_2024}. Finally, it is possible for different levels of extinction for the blue- and red-shifted lobes to contribute to observed asymmetries \citep{Dutta_2025}; in extreme cases, different levels of extinction can lead to jets appearing monopolar \citep[e.g.][]{Harsono_2023, Dutta_2024}.

\subsection{Jet Velocities and Velocity Variations}\label{sec:velocity_discussion}

The P-V diagrams in Figure~\ref{fig:pv_fits} show that the jets of HOPS~153 and HOPS~370, both at a distance of 390~pc, have relatively constant velocities along their entire jets. The blue-shifted lobe of HOPS~153 has an average $V_{rad}$ of -46$~\pm~4~km~s^{-1}$ from 50 to 1300~au, and the red-shifted lobe has an average $V_{rad}$ of +43$~\pm~4~km~s^{-1}$ from -300 to -1300~au. The blue-shifted lobe of HOPS~370 has an average $V_{rad}$ of -37$~\pm~3~km~s^{-1}$ from 50 to 1300~au, and the red-shifted lobe has an average $V_{rad}$ of +40$~\pm~4~km~s^{-1}$ from -1300 to -300~au.

There is a slope in the P-V diagrams of HOPS~153 and HOPS~370 at the interface of the blue- and red-shifted lobes, but these slopes are spatially coincident with the regions of highest extinction near the protostellar disks (Figures~\ref{fig:hops153_jetmaps}, \ref{fig:hops370_jetmaps}), and only cover $<$10 pixels ($<$5 resolution elements); therefore at the distance of the protostars the velocities of the red- and blue-shifted lobes are blended in the inner $\sim$5 pixels. For B335, at a distance of 165~pc, the slope in the P-V diagram is spatially resolved, out to a distance of 330~au for the blue-shifted lobe and 470 au for the red-shifted lobe. This is not simply an effect of the smaller distance to B335 (165~pc) relative to HOPS~153 and HOPS~370 (390~au); the velocities of the jet from the low-mass IRAS~16253, at a distance of 140~pc, was observed to remain constant along the length of the inner jet, similar to HOPS~153 and HOPS~370. The slope is roughly symmetrical and relatively constant, with a value of 0.16~km~s$^{-1}$~au$^{-1}$ for the blue-shifted lobe and a value of 0.12~km~s$^{-1}$~au$^{-1}$ for the red-shifted lobe. MHD models predict the magnetic lever arm accelerates collimated jets to terminal velocity within a distance of 100 times the launching radius in the inner disk \citep[e.g.][]{Tabone_2020, Lee_2022}; estimating a Keplerian launch radius of $\sim$0.02~au corresponding to a total velocity of 166~km~s$^{-1}$, the observed change in velocity out to a few hundred au is too far from the central protostar for magnetic acceleration ($<$20~au). 

Given the inclination of the B335 jet, we cannot discount the possibility that the observed gradient in velocity with distance is due in part to a blending of the velocities from the blue- and red-shifted lobes in the same beam, or from (symmetrical) precession of the jet in the line of sight. If we assume the jet is actually moving in the line of sight with a constant deprojected velocity over the length of the jet, the maximum radial velocity (for which we adopt the maximum in the blue-shifted lobe at 330~au), when corrected for the inclination derived from the $V_{rad}$ and $V_{tan}$, should be equal to the minimum velocity corrected for the inclination angle at launch (Equation~\ref{eq1}):
\begin{equation}\label{eq1}
\frac{V_{rad|330~au}}{cos(\theta_{1})} = \frac{V_{rad|66~au}}{cos(\theta_{2})} = 166~km~s^{-1};
\end{equation}
where $V_{rad|330~au}=90~km~s^{-1}$ is the radial velocity of the jet at 330~au and $\theta_{1}=57^{\circ}$ is the inclination angle of the jet at 330~au derived from the proper motion of the shocked knot B335-C (Section~\ref{sec:inclination}). 
Taking the $V_{rad}$ two spatial resolution elements from the protostar position as the minimum velocity (41~$\pm$1~km~s$^{-1}$ at 66~au for the 17.936~$\mu$m [Fe~II] line, uncorrected for inclination), the launch angle required to reproduce the observed gradient in the $V_{rad}$ is 76$^{\circ}$. This implies that if the jet velocity is constant and the observed gradient is due to motion along the line of sight, this motion must be on the order of 76$^{\circ}$ - 57$^{\circ}$ = 19$^{\circ}$. If, instead, we assume the inclination angle at launch is the 83.5$^{\circ}$ listed in Table~\ref{tab:sources}, this would require an uncorrected $V_{rad}$ of 19~km~s$^{-1}$ under the assumption of a constant total jet velocity. Given the 41~$\pm$1~km~s$^{-1}$ $V_{rad}$ at the base of the jet estimated from the 17.9 [Fe II] line, even accounting for the relative wavelength calibration uncertainty in Channel 3 Long of $\pm$6~km~s$^{-1}$ \citep{Argyriou_2023,Pontoppidan_2024b,Banzatti_2025}, a larger magnitude line-of-sight precession of 26.5$^{\circ}$ from the 83.5$^{\circ}$ inclination of the outflow cavity is unlikely. The apparent difference between the cavity inclination from radiative transfer modeling and the jet inclination from the radial and tangential velocity components is plausibly due to precession of the jet, causing the observed velocity gradient in the P-V diagram.

\begin{figure*}[t]
    \centering
    \includegraphics[scale=0.5]{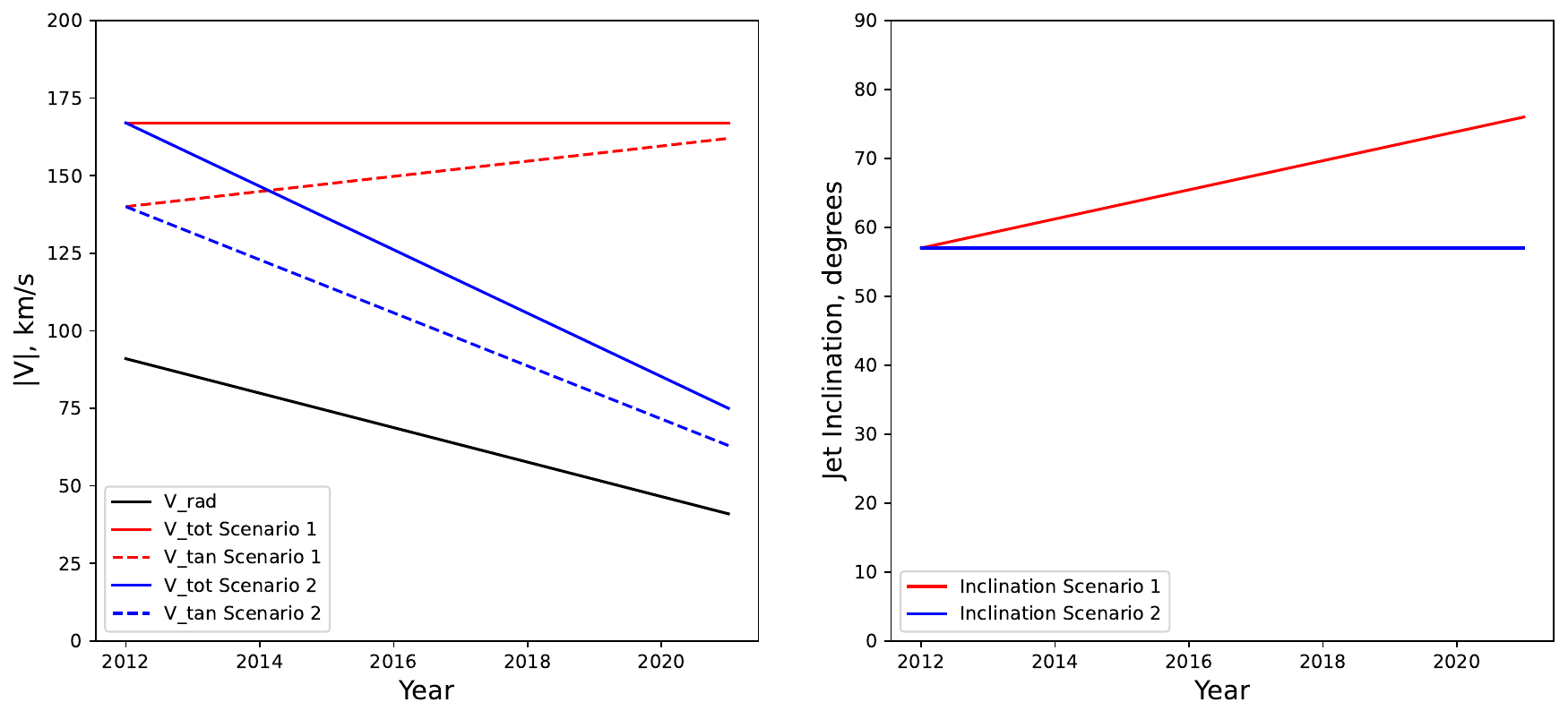}
    \includegraphics[scale=0.5]{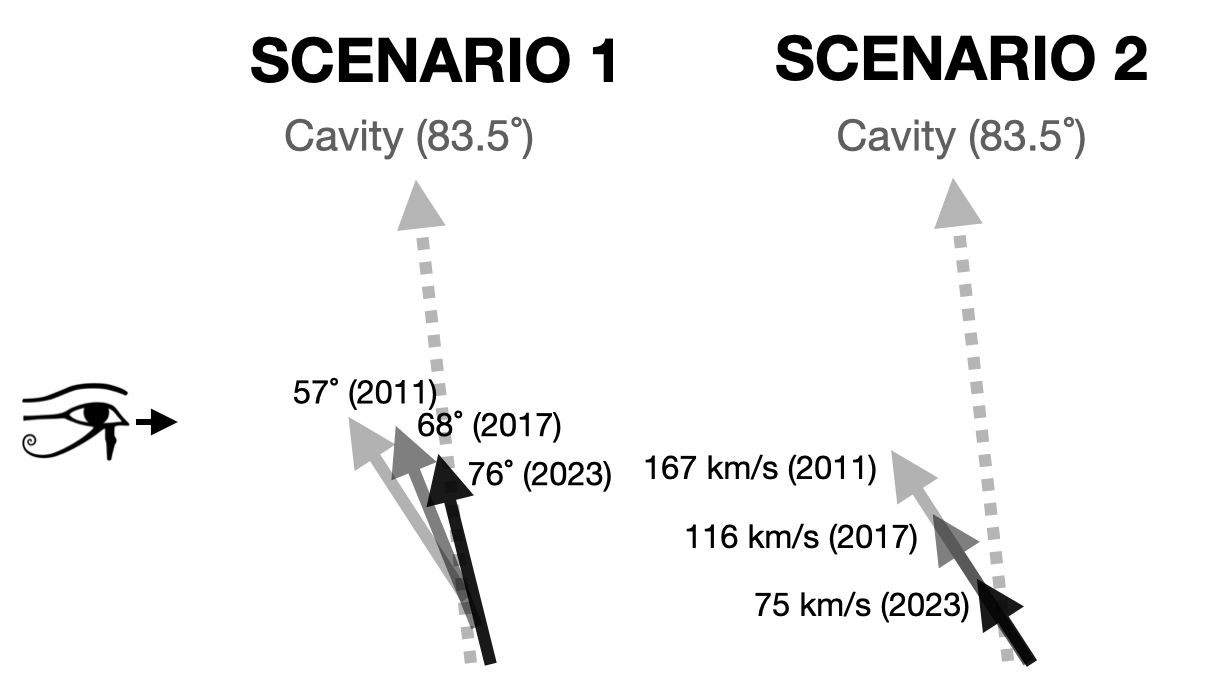}
    \caption{A demonstration of the two scenarios to explain the symmetrical change in radial velocity of the B335 jet with time. Scenario 1, shown in red, has a constant total (i.e. deprojected) launch velocity with time, with the inclination angle of the jet varying. Scenario 2, shown in blue, instead has a constant jet inclination angle in time with a decreasing launch velocity. These scenarios are demonstrated by the cartoons in the second row.}
    \label{fig:scenarios}
\end{figure*}

However, it is equally plausible that the jet launching velocity is not constant with time, in which case the P-V diagram of B335 would indicate that at an earlier epoch the jet was being launched with a higher velocity than the present. Taking the 140~km~s$^{-1}$ tangential jet velocity at a distance of 330~au, the estimated date for the end of the higher-velocity epoch is 2011 (12 years prior to the JWST observations). From that point, in this scenario, the jet launching velocity decreases at an average rate of 8.25~km~s$^{-1}$~yr$^{-1}$ until today, when the current observable total jet launching velocity is 75~$\pm$2~km~s$^{-1}$ (the $V_{tan}$ corresponding to $V_{rad}$= 41~$\pm$1~km~s$^{-1}$, assuming a constant inclination of 57$^{\circ}$). If this were the case, it would indicate that the Keplerian launching radius of the collimated jet moved radially outward from 0.1~au to 0.4~au over the course of 12 years. These two scenarios are illustrated in Figure~\ref{fig:scenarios}, and can be tested through future observations of proper motions of the shocked knots in B335, as the different scenarios imply different tangential velocities over time.

It is possible that there is a connection between jet velocity and protostellar mass accretion rate, but there is scant observational evidence to date that supports this hypothesis \citep[for some possible examples, see][]{Jhan_2022, Dutta_2024, Takahashi_2024}. Tantalizingly, in 2023 it was reported that B335 had completed a ~10 year outburst with an increase in NIR luminosity by a factor of 5-7 \citep{Evans_2023}. 
It is unclear if or how this outburst event is connected to the launch of the shocked knot and the possible decrease in jet launching velocity over the same time span, and searching for correlations from a larger sample is required to show a correlation between variability and outflow rates. For a possible connection between the accretion burst in B335 and shocked knots in the high-velocity CO outflow observed with ALMA, see \citet{Kim_2024}. We discuss the evidence supporting a connection in more detail in Section~\ref{sec:b335_lc}.

\subsubsection{Inclination of the B335 Cavity and Jet} \label{sec:b335_inc}

The jet of B335 was shown in Section~\ref{sec:inclination} to have a quite different inclination angle compared to that of the outflow cavity determined from radiative transfer modeling of the scattered light from the cavity walls. This inspired a deeper investigation of the inclination of this source and its outflow. The inclination angle of the large-scale outflow was determined to be $\sim80^{\circ}$ by \citet{Hirano_1988}. The inclination determined from the SED was estimated at $\sim 87^{\circ}$ by \citet{Stutz_2008}, by comparing to the grid of radiative transfer model SEDs from \citet{Robitaille_2006, Robitaille_2007}. This angle was adopted by \citet{Evans_2023} without further testing to develop a more complete radiative transfer model of the source, constrained by many more observations. 

Given the inconsistency of this inclination angle ($\theta_{\rm cavity}$) with that of the jet ($\theta_{\rm jet}$), we tested other cavity inclination angles versus the same observational data, using the same modeling method and other source parameters used by \citet{Evans_2023}. Interestingly, the 83.5$^{\circ}$ average of the two inclination angles derived above is the best fit to the JWST data in the resulting radiative transfer models. When the inclination angle is decreased below $83^{\circ}$, the model rapidly begins to over-predict the IRAC flux in the $2\farcs4$ aperture from \citet{Stutz_2008}. For $\theta_{\rm cavity} = 57^{\circ}$ (i.e. the inclination angle determined from V$_{rad}$ and the V$_{tan}$ of the shocked knot B335-C), the prediction is a factor of 40 to 48 larger than the observations for the three IRAC bands from 4.5 to 8.0 \micron, which are reasonably well-modeled at $\theta_{\rm cavity} > 83^{\circ}$. The inconsistency between the inclination angle of the large-scale cavity and the ionized jet strongly indicates a change with time, as would happen if the jet precesses along the line of sight.

\subsection{Association of B335 NIR Variability with Jet Morphology and Kinematics}\label{sec:b335_lc}

\begin{figure}[h]
    \centering
    \includegraphics[scale=0.6]{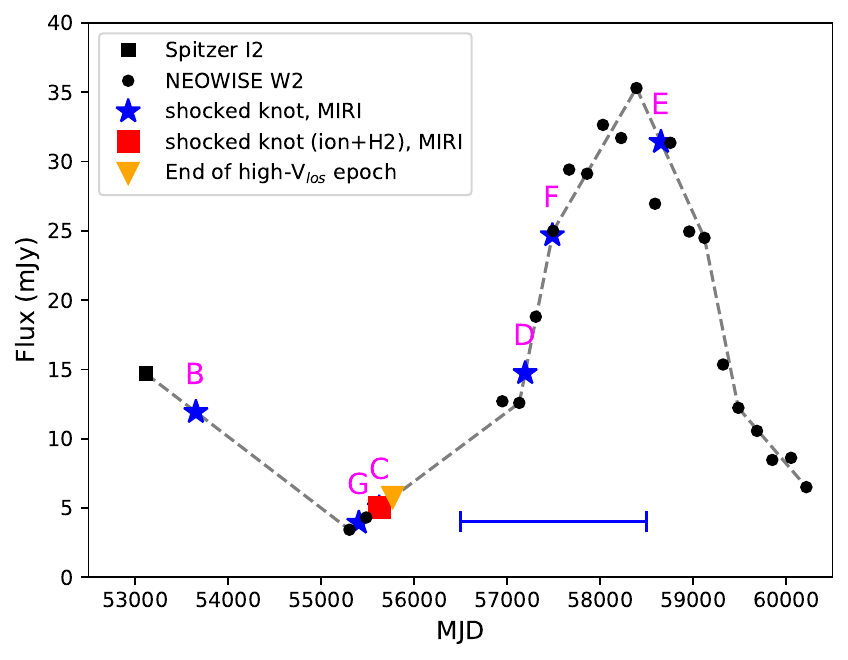}
    \caption{The NIR light curve of B335 over the past two decades, as observed with Spitzer and NEOWISE. Black circles mark NEOWISE W2 photometric points, and the black square marks a Spitzer photometric point at MJD 53116.3, with a different sized aperture. Black dashed lines show linear fits to pairs of points in the light curve, marking out 7 distinct epochs with different slopes; however, the sparse sampling means that the variability is unconstrained where there are no photometric points. Stars mark the launch dates of shocked knots identified in MIRI, using the fits to determine their y-axis positions; the red square marks the position of a shocked knot observed in molecular, atomic, and ionic tracers as detailed in \citet{Federman_2024}. The uncertainty in the dynamical time of the shocked knots is shown at the bottom of the figure. The yellow triangle marks the end of the high-V$_{rad}$ epoch noted in Figure~\ref{fig:pv_fits} and Section~\ref{sec:velocity_discussion}; the uncertainty in this date is the same as for the shocked knots.}
    \label{fig:b335_lc}
\end{figure}

In Figure~\ref{fig:b335_lc} we display the NIR light curve of B335 as observed with Spitzer I2 and NEOWISE W2, starting in 2003 (MJD 53116.3) and covering two decades \citep{Evans_2023, Kim_2024, Lee_2025}. The photometric data making up the light curve are displayed with black circles for NEOWISE and a black square for the one Spitzer epoch in 2003. We then fit lines to pairs of photometric points with dotted black lines. For all the points and their corresponding dates in the figure, other than the photometric data, the y-axis values are interpolated along these lines, but do not represent actual photometric fluxes of the knots in the jet of B335. They do not take into account variability between the measurements, which is unconstrained by the available data.

We measure the dynamical timescales of the shocked knots identified in the MIRI MRS data (observed in 2023, MJD 60077.5; 
Appendix~\ref{sec:knot_regions}) using Gaussian fits to determine their positions (Table~\ref{tab:knot_pos}) and the tangential velocity of 140~km~s$^{-1}~\pm$~28~km~s$^{-1}$ as detailed in Section~\ref{sec:inclination}, assuming this velocity is constant for all the knots. The approximate ejection dates for the knots, determined from the dynamical timescales, are shown with blue stars. The uncertainty in the dates determined using this tangential velocity is $\pm1000$~days. We use the same tangential velocity to determine the dynamical timescale and date for the end of the high-V$_{rad}$ epoch noted in Section~\ref{sec:velocity_discussion} and Figure~\ref{fig:pv_fits} (yellow triangle in the figure). \citet{Hodapp_2024} identified two of the shocked knots in higher-resolution NIRCam data, along with two others closer to the driving source (i.e. at later epochs), and determined their approximate ejection dates; the NIRCam data, however, primarily traces molecular emission.

In Figure~\ref{fig:b335_lc} it can be seen that sometime between MJD 55304.2 and 56948.5 the 4.5~$\mu$m flux began to rise; the proximity of a photometric point at MJD 55485.0 to the ejection of two shocked knots (B335-C and B335-G), one of which (B335-C) is observed in molecular emission, suggests that these knots were ejected within three years before or after the beginning of the rise in flux. The end of the high-V$_{rad}$ epoch occurred 1$\pm$3~yr later, when the V$_{rad}$ began to decrease steadily due to precession in the line of sight and/or a decrease in total launch velocity. Beginning at MJD 57130.9, roughly coincident with the ejection of another shocked knot bright in [Ar~II] and [Ne~II] (knot B335-D), the flux began to rise with a much sharper slope, the beginning of the ''burst" proper. This lasted until MJD 57495.1 when the slope in the light curve became shallower. The flux decreased from the peak of the burst until MJD 59121.9, when the slope of the decrease became steeper. The shallower decrease continued until the most recent NEOWISE observation at MJD 60217.3. 

Due to the sparse sampling, it is difficult to draw conclusions regarding a connection between NIR variability \citep[often attributed to accretion variability, e.g.][]{Jhan_2022, Dutta_2024, Takahashi_2024, Fischer_2023}, and changes in the morphology and kinematics of the ionized jet. There is apparently a higher frequency of shocked knots during the burst event relative to the pre-burst epoch, with five knots located in the elevated flux phase of the light curve versus the one knot (B335-B) in the epoch between the Spitzer photometry and the first NEOWISE point. However, given the sparse sampling of the photometry and the small number of knots, this evidence remains tentative and circumstantial. Regular monitoring of the protostar and its outflow is required to test this hypothesis.

\section{Conclusions}

We analyzed the morphology and kinematics of the shock-ionized jets from Class~0 protostars B335 (mass of central protostar: 0.25~M$_{\odot}$), HOPS~153 (0.6~M$_{\odot}$), and HOPS~370 (2.5~M$_{\odot}$), using JWST NIRSpec and MIRI MRS IFU data from the IPA program. We examined the jets and found evidence of precession, deflections, and deviations from bipolar symmetry as seen in the shock-ionized [Fe II] emission. We found evidence of the jets getting wider with distance from the protostar. Finally, we examined the kinematics of the jets, determining the total [Fe II] jet velocities for the three protostars. Below we enumerate the conclusions of the analysis of the collimated, ionized jets of B335, HOPS~153, and HOPS~370:

\begin{enumerate}

\item We identified shocked knots of ionic line emission in the jets of the three protostars. The number and projected distance of these knots are not symmetric between blue-shifted and red-shifted lobes of the jets. 

\item All three jets, despite significant differences in environment, show wiggles along their length with angular deflections ranging from 2.3$^{\circ}$ to 13.8$^{\circ}$. Such wiggles are asymmetric, appearing in only one lobe of each of the three protostellar jets. There are also larger deflections with amplitudes $>$40~au at farther distances from the protostar along the jet. 

\item The angle between the blue- and red-shifted lobes of B335 is 168.5$^{\circ}$~$\pm$~0.3$^{\circ}$, an 11.5$^{\circ}$~$\pm$~0.3$^{\circ}$ deviation from 180$^{\circ}$ bipolar symmetry in launch angle of the lobes. In contrast, HOPS~153 and HOPS~370 do not show significant deviation from 180$^{\circ}$ between blue- and red-shifted lobes. 

\item The widths across the jets have complicated structures, with significant variations along the length of the jets. Nevertheless, there is a trend for the red-shifted lobes of B335 and HOPS~370 to widen with distance. The opening angles of all the other jet lobes can only be constrained by upper limits. The opening angles of the inner jets within 1500~au range from $2.1^{\circ}$ to $<10^{\circ}$, corresponding to average radial jet expansion velocities of 5.1 to $<21.3~km~s^{-1}$. 

\item We determined the radial velocity structure along the jet from the position-velocity diagrams of the shocked [Fe~II] emission lines. We corrected the radial velocities for the inclination of the protostellar disks, with total jet velocities of 184 and 154~km~s$^{-1}$ for HOPS~153 and HOPS~370, respectively.

\item HOPS~153 and HOPS~370 show relatively constant velocities along the lengths of their jets. In contrast, there is a clearly resolved slope in the position-velocity diagram for B335 which we attribute to precession of the jet along the line of sight.

\item Correcting the radial velocity of the B335 jet for the 83.5$^{\circ}$ inclination determined from modeling of the scattered light from the walls of the outflow cavity and the large-scale outflow results in an unrealistically high total jet velocity of 800~km~s$^{-1}$. Due to the proximity of B335 (165~pc) and a fortuitous 8 month gap in time between the observations of the NIRSpec and MIRI data, we measured the proper motion of a shocked knot to determine the tangential velocity along the jet. This, combined with the radial velocity, results in a realistic total jet velocity of 166~km~s$^{-1}$ with an inclination angle of 57$^{\circ}$. We hypothesize that the discrepancy in inclination angles between the jet and the outflow cavity is due to precession of the jet along the line of sight. 

\item There is no correlation with the dynamical times of knots in the jet of B335 with its outburst. However, there is a suggestion that the frequency of knots increased during the burst, and a correlation with the presence of molecules and bright argon and neon emission in the knots during the burst.

\item The observed asymmetric wiggles and bends in the jets are likely due to interaction of the circumstellar disk with the stellar magnetic field and/or the larger-scale magnetic field threaded through the disk causing warps, changing the direction of the angular momentum vector of the inner disk, and therefore the jet launching vector. Disk warping may also be caused by asymmetric infall from streamers onto the disks. Observed asymmetries in shocked knots, wiggles, jet widths, and launch angles can also be attributed to interactions of the disk with the magnetic field(s).


\end{enumerate}

\begin{acknowledgments}

\noindent
This work is based on observations made with the NASA/ESA/CSA James Webb Space Telescope. The data were obtained from the Mikulski Archive for Space Telescopes at the Space Telescope Science Institute, which is operated by the Association of Universities for Research in Astronomy, Inc., under NASA contract NAS 5-03127 for JWST. These observations are associated with program \#1802. All the JWST data used in this paper can be found in MAST: \dataset[10.17909/3kky-t040]{http://dx.doi.org/10.17909/3kky-t040}. 

\noindent
Support for SF, AER, STM, RG, WF, JG, JJT and DW in program \#1802 was provided by NASA through a grant from the Space Telescope Science Institute, which is operated by the Association of Universities for Research in Astronomy, Inc., under NASA contract NAS 5-03127.

\noindent
SF and ACG acknowledge support from PRIN-MUR 2022 20228JPA3A “The path to star and planet formation in the JWST era (PATH)” funded by NextGeneration EU and by INAF-GoG 2022 “NIR-dark Accretion Outbursts in Massive Young stellar objects (NAOMY)” and Large Grant INAF 2022 “YSOs Outflows, Disks and Accretion: towards a global framework for the evolution of planet forming systems (YODA)”. 

\noindent
NJE thanks the University of Texas at Austin for research support.

\noindent
AS gratefully acknowledges support by the Fondecyt Regular (project code 1220610), and ANID BASAL project FB210003.

\noindent
G.A. and M.O. acknowledge financial support from grants PID2020-114461GB-I00 and CEX2021-001131- S, funded by MCIN/AEI/10.13039/501100011033.

\noindent
P.N. acknowledges support from the ESO and IAU Gruber Foundations Fellowships.

\noindent
RK acknowledges financial support via the Heisenberg Research Grant funded by the Deutsche Forschungsgemeinschaft (DFG, German Research Foundation) under grant no.~KU 2849/9, project no.~445783058.

\noindent The work by MN was carried out at the Jet Propulsion
Laboratory, California Institute of Technology, under a contract with the National Aeronautics and Space Administration (80NM0018D0004).

\noindent AS gratefully acknowledges support by the Fondecyt Regular (project code 1220610), ANID BASAL project FB210003, and the China-Chile Joint Research Fund (CCJRF No. 2312).

\noindent CK acknowledges support from the European Union (ERC Starting Grant
DiscEvol, project number 101039651) and from Fondazione Cariplo, grant
No. 2022-1217. Views and opinions expressed are, however, those of the
author(s) only and do not necessarily reflect those of the European
Union or the European Research Council. Neither the European Union nor
the granting authority can be held responsible for them.

\noindent
We gratefully acknowledge valuable conversations with Dr. Sylvie Cabrit and Dr. Tom Ray.

\end{acknowledgments}

%






\appendix

\section{Precision in Gaussian Fitting to [Fe~II] Data}\label{sec:gauss_precision}

In Table~\ref{tab:gauss_precision} we list the average limits of precision in our Gaussian fitting to the spatial and kinematic data, where the precision is defined as resolution/SNR. We use the SNR at the pixel where the value is closest to the mean of the entire FOV at the wavelength of interest. We only consider pixels where the SNR~$>$~5.

\begin{table}[h]
\caption{Limits of the precision in Gaussian fitting to spatial and kinematic data for each of the [Fe~II] lines used in the analysis. Only data points with SNR~$>$~5 are considered.}
\centering
\begin{tabular}{ccccc}
    [Fe~II] Line & Avg. SNR & Angular Precision ('')  & Spatial Precision (au)  & Velocity Precision (km~s$^{-1}$)  \\ \hline \hline
    \multicolumn{5}{c}{\textbf{B335}} \\
    4.115~$\mu$m & 6 & 0.03 & 5.5 & 53.0 \\ \hline 
    5.340~$\mu$m & 7 & 0.04 & 6.6 & 12.1 \\ \hline 
    17.936~$\mu$m & 73 & 0.01 & 1.6 & 1.2 \\ \hline 
    24.519~$\mu$m & 14 & 0.07 & 11.0 & 11.8 \\ \hline 
    25.988~$\mu$m & 32 & 0.03 & 4.9 & 4.3 \\ \hline \hline
    \multicolumn{5}{c}{\textbf{HOPS~153}} \\ 
    4.115~$\mu$m & 5 & 0.04 & 14.6 & 58.9 \\ \hline 
    5.340~$\mu$m & 5 & 0.06 & 21.3 & 16.6 \\ \hline 
    17.936~$\mu$m & 6 & 0.11 & 42.7 & 13.9 \\ \hline 
    24.519~$\mu$m & 5 & 0.17 & 66.0 & 30.0 \\ \hline 
    25.988~$\mu$m & 7 & 0.14 & 54.0 & 20.0 \\ \hline \hline
    \multicolumn{5}{c}{\textbf{HOPS~370}} \\
    4.115~$\mu$m & 5 & 0.04 & 14.3 & 57.9 \\ \hline 
    5.340~$\mu$m & 13 & 0.02 & 8.3 & 6.4 \\ \hline 
    17.936~$\mu$m & 40 & 0.02 & 6.8 & 2.2 \\ \hline 
    24.519~$\mu$m & 6 & 0.14 & 55.4 & 25.2 \\ \hline 
    25.988~$\mu$m & 58 & 0.02 & 6.5 & 2.4 
\end{tabular}
\tablecomments{For the SNR column, values are listed as maximum/mean SNR. For the angular, spatial, and velocity precision columns, values are listed as minimum/mean.}
\label{tab:gauss_precision}
\end{table}

\section{Gaussian Fitting to Knots}\label{sec:knot_regions}
\begin{figure}[h]
    \centering
    \includegraphics[width=\textwidth]{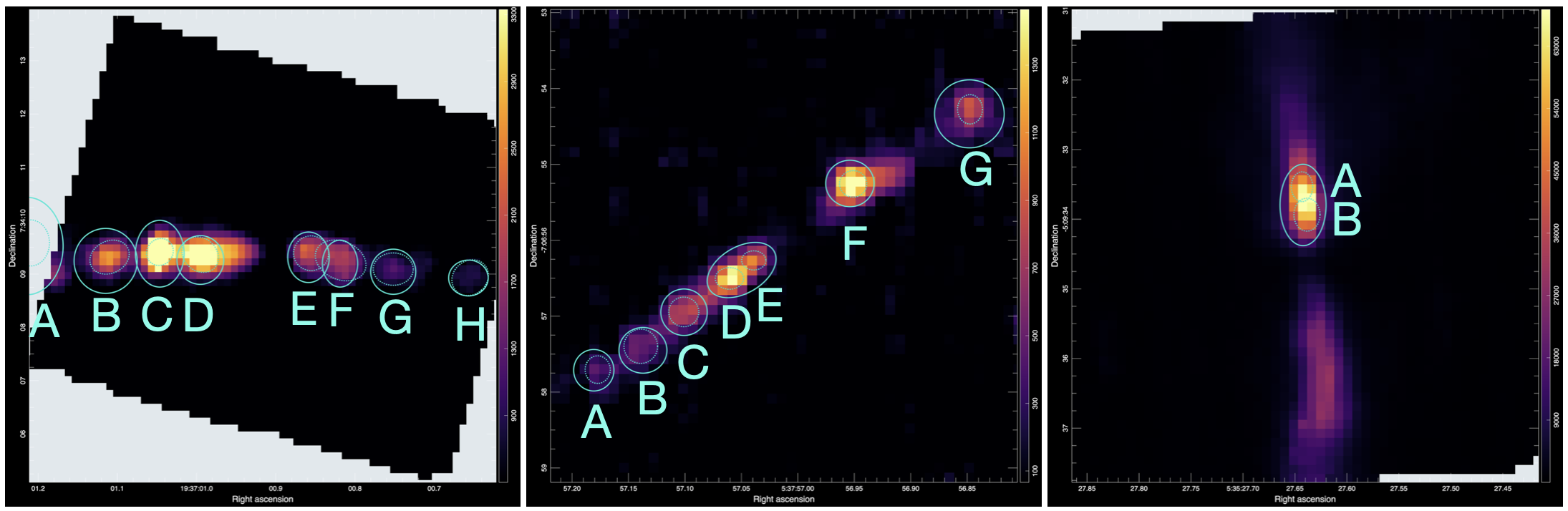}
    \caption{[Fe~II] 5.340~$\mu$m line maps used to identify shocked knots in the jets by eye, for B335, HOPS~153, and HOPS~370 (left to right, respectively). Cyan circles with solid boundaries show the regions over which the fits were applied, and the ellipses with dashed boundaries show the best fits. There is an additional knot at the eastern edge of the FOV for B335 that was identified using the [Fe~II] 17.936~$\mu$m line.}
    \label{fig:knot_pos}
\end{figure}

In Figure~\ref{fig:knot_pos} we demonstrate the method used to identify shocked knots of emission in the [Fe~II] 5.340~$\mu$m line maps (cataloged in Tables~\ref{tab:knot_pos}). We fit 2D Gaussian profiles within the regions shown in the figure to determine the centers of the knots (this intrinsically assumes the knots are Gaussian in shape). For this purpose, we used the CARTA analysis software \citep{Comrie_2021}. In Table~\ref{tab:knot_size} we list the Gaussian FWHM sizes of the knots determined from the fits.

\begin{table}[h]
\caption{FWHM sizes of shocked knots identifed in the [Fe~II] MIRI MRS data, from 2D Gaussian fits. The fits are from the [Fe~II] 5.340~$\mu$m line, except knot B335-A which falls outside the FOV of the 5.340~$\mu$m line and was instead measured with the 17.936~$\mu$m line. The final column gives the ratio of the FWHM of the Gaussian fits to the FWHM$_{PSF}$ from \citet{Law_2023}.}
\centering
\begin{tabular}{cccc}
    Knot & FWHM & FWHM$_{uncertainty}$ & FWHM$_{fit}$/FWHM$_{PSF}$\\ \hline \hline
    \multicolumn{4}{c}{\textbf{B335}} \\
    A* & 0\farcs86~$\times$~0\farcs70 & $\pm$~0\farcs11~$\times$~$\pm$~0\farcs10 & 1.2~$\times$~1.0 \\ \hline
    B & 0\farcs60~$\times$~0\farcs77 & $\pm$~0\farcs05~$\times$~$\pm$~0\farcs06 & 2.1~$\times$~2.8 \\ \hline
    C & 0\farcs48~$\times$~0\farcs53 & $\pm$~0\farcs05~$\times$~$\pm$~0\farcs05 & 1.7~$\times$~1.9 \\ \hline
    D & 0\farcs73~$\times$~0\farcs61 & $\pm$~0\farcs06~$\times$~$\pm$~0\farcs05 & 2.6~$\times$~2.2 \\ \hline
    E & 0\farcs65~$\times$~0\farcs66 & $\pm$~0\farcs06~$\times$~$\pm$~0\farcs06 & 2.3~$\times$~2.4 \\ \hline
    F & 0\farcs99~$\times$~0\farcs72 & $\pm$~0\farcs05~$\times$~$\pm$~0\farcs03 & 3.5~$\times$~2.6 \\ \hline
    G & 0\farcs61~$\times$~0\farcs75 & $\pm$~0\farcs04~$\times$~$\pm$~0\farcs05 & 2.2~$\times$~2.7 \\ \hline
    H & 0\farcs72~$\times$~0\farcs63 & $\pm$~0\farcs05~$\times$~$\pm$~0\farcs04 & 2.6~$\times$~2.3 \\ \hline \hline
    \multicolumn{4}{c}{\textbf{HOPS~153}} \\ 
    A & 0\farcs36~$\times$~0\farcs33 & $\pm$~0\farcs11~$\times$~$\pm$~0\farcs10 & 1.3~$\times$~1.2 \\ \hline
    B & 0\farcs45~$\times$~0\farcs44 & $\pm$~0\farcs08~$\times$~$\pm$~0\farcs08 & 1.6~$\times$~1.6 \\ \hline
    C & 0\farcs39~$\times$~0\farcs40 & $\pm$~0\farcs06~$\times$~$\pm$~0\farcs06 & 1.4~$\times$~1.4 \\ \hline
    D & 0\farcs29~$\times$~0\farcs37 & $\pm$~0\farcs05~$\times$~$\pm$~0\farcs06 & 1.0~$\times$~1.3 \\ \hline
    E & 0\farcs25~$\times$~0\farcs32 & $\pm$~0\farcs08~$\times$~$\pm$~0\farcs10 & 0.9~$\times$~1.1 \\ \hline
    F & 0\farcs35~$\times$~0\farcs36 & $\pm$~0\farcs06~$\times$~$\pm$~0\farcs06 & 1.3~$\times$~1.3 \\ \hline
    G & 0\farcs39~$\times$~0\farcs33 & $\pm$~0\farcs04~$\times$~$\pm$~0\farcs03 & 1.4~$\times$~1.2 \\ \hline \hline
    \multicolumn{4}{c}{\textbf{HOPS~370}} \\
    A & 0\farcs43~$\times$~0\farcs31 & $\pm$~0\farcs11~$\times$~$\pm$~0\farcs08 & 1.5~$\times$~1.1 \\ \hline
    B & 0\farcs47~$\times$~0\farcs38 & $\pm$~0\farcs11~$\times$~$\pm$~0\farcs09 & 1.7~$\times$~1.4 \\
\end{tabular}
\tablecomments{Knot B335-A falls outside of the FOV of the 5.340~$\mu$m data and was fit using the 17.936~$\mu$m data.}
\label{tab:knot_size}
\end{table}

\section{Absolute Velocity Offsets}\label{sec:vel_offsets}
\begin{figure}[h]
    \centering
    \includegraphics[width=0.9\textwidth]{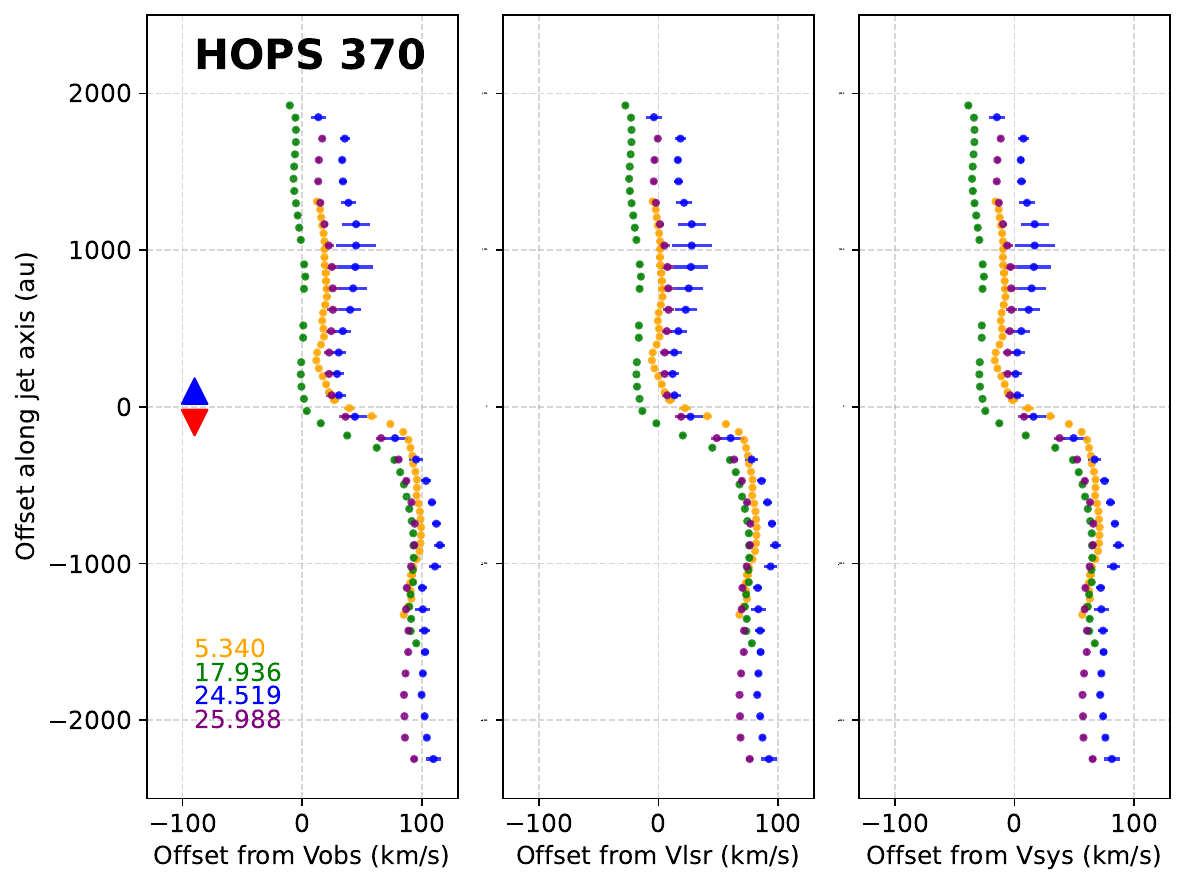}
    \caption{The centers of the Gaussian fits to the P-V diagrams for each of the [Fe~II] lines accessible by MIRI for HOPS~370. From left to right is the observed velocity shift relative to the barycentric reference frame, LSR, and the systemic velocity. The lines are all strongly redshifted, indicating an unaccounted uncertainty in the absolute radial velocities. Positive y-axis values correspond to the blue-shifted lobes, and negative y-axis values correspond to the red-shifted lobes.}
    \label{fig:vel_offsets}
\end{figure}

As mentioned in Section~\ref{sec:kinematics}, we chose to assume symmetric radial velocities due to non-physical results when using absolute velocity shifts relative to the Local Standard of Rest (LSR) or the systemic velocity. In Figure~\ref{fig:vel_offsets}, we show the results of the [Fe~II] P-V diagrams for HOPS~370 when we do not assume symmetric velocities. The observed data is set to the barycentric reference frame during the data reduction process. We corrected for LSR using the relationship from \citet{Schonrich_2018}, and corrected for the system velocity of +11.2~km~s$^{-1}$ from \citet{Tobin_2019}.

As can be seen in the figure, whether in the barycentric, LSR, or systemic reference frame, the velocities of the iron lines are strongly red-shifted, with the ostensibly blue-shifted lobe centered around 0~km~s$^{-1}$ and the red-shifted lobe between 50 and 100~km~s$^{-1}$. Although we might expect that the velocities between blue and red-shifted lobes are not perfectly symmetrical, such a large discrepancy is clearly non-physical. It is not clear why this shift is present, or whether it is due to a calibration or data reduction error. Regardless, it indicates a large uncertainty in the absolute velocities of the blue and red-shifted lobes, for which reason we adopt the assumption of symmetrical velocities by subtracting out the mean as was done in \citet{Narang_2024} and detailed in Section~\ref{sec:kinematics}.

\section{HOPS~370 Jet in 5~cm Radio Continuum }\label{sec:hops370_vla}

\begin{figure}[hb]
    \centering    
    \includegraphics[width=0.8\textwidth]{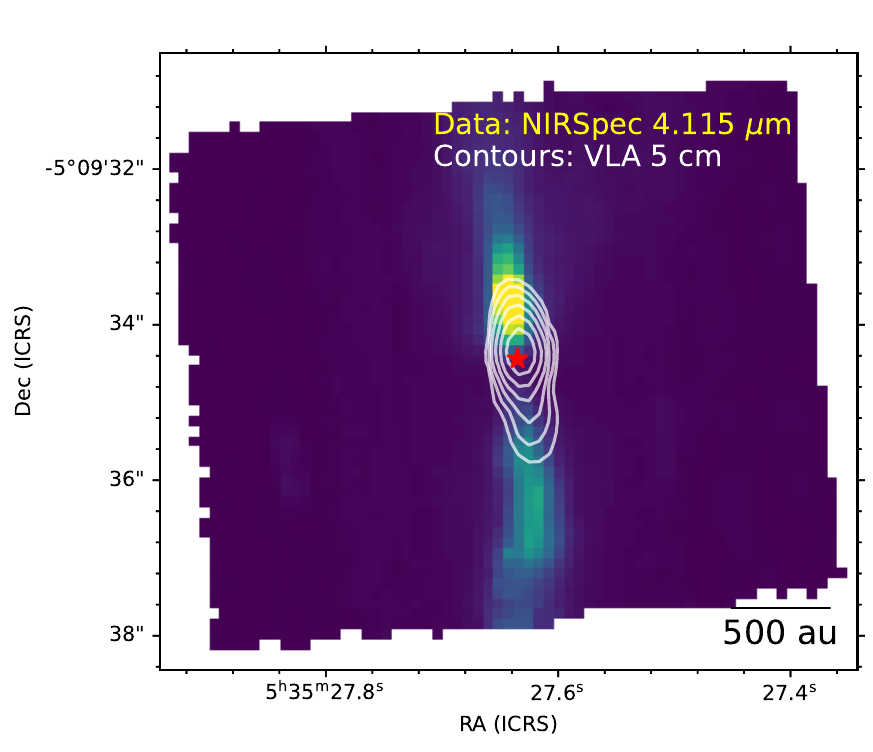}
    \caption{MIRI MRS [Fe~II] 5.340~$\mu$m data of HOPS~370 with the 5~cm VLA contours \citep[from][]{Osorio_2017} overlaid in white. The ALMA position of HOPS~370 is marked by a red star. The radio continuum jet has a very similar position angle to the IR [Fe~II] jet.}
    \label{fig:hops370_vla}
\end{figure}

In Figure~\ref{fig:hops370_vla} we show the MIRI [Fe~II] 5.340~$\mu$m data of HOPS~370 with the 5~cm VLA contours \citep{Osorio_2017} overlaid. This demonstrates the consistency in position angle between the cm radio continuum and the IR [Fe~II] jets. There was 8 years separating the MIRI and VLA observations.

\clearpage
\bibliography{main_bib-1}{}
\bibliographystyle{aasjournal}



\end{document}